\documentclass[useAMS,usenatbib,usegraphicx,usedcolumn]{mn2e}
\usepackage{fixltx2e}
\usepackage{amsmath}
\usepackage{mathrsfs}

\newcommand{\betam}{\beta_{\rm ||}}
\newcommand{\betan}{\beta_{\rm \perp}}

\title[AGN Coronal Emission models I.]{AGN Coronal Emission models I. The Predicted Radio Emission}

\author[I.~Raginski, and A.~Laor]
{I.~Raginski,$^1$\thanks{E-mail: raginski.igor@gmail.com} and Ari Laor$^1$  \\
$^1$Physics Department, Technion -- Israel Institute of Technology, Haifa~32000, Israel}

\begin{document}
	\date{}
	\pagerange{\pageref{firstpage}--\pageref{lastpage}} \pubyear{2015}
	\maketitle

	\label{firstpage}

\begin{abstract}
Accretion discs in AGN may be associated with coronal gas, as suggested by their X-ray
emission. Stellar coronal emission includes radio emission,
and AGN corona may also be a significant source for radio emission in radio quiet (RQ) AGN. We calculate the coronal
properties required to produce the observed radio emission in RQ AGN, either from synchrotron emission
of power-law (PL) electrons, or from cyclo-synchrotron emission of hot mildly-relativistic 
thermal electrons. We find that a flat spectrum, as observed in about half of RQ AGN, can be produced by
corona with a disc or a spherical configuration, which extends from the innermost regions out to a pc scale.
A spectral break to an optically thin power-law emission is expected around 300-1000~GHz, 
as the innermost corona becomes optically thin. In case of thermal electrons, a sharp spectral cutoff
is expected above the break. The position of the break can be measured with VLBI observations,
which exclude the cold dust emission, and it can be used to probe the properties of the
innermost corona. Assuming equipartition of the coronal
thermal energy density, the PL electrons energy density, and the magnetic field, we find that the
energy density in a disc corona should scale as $\sim R^{-1.3}$, to get a flat spectrum. In the
spherical case the energy density scales as $\sim R^{-2}$, and is 
$\sim 4\times 10^{-4}$
of the AGN radiation energy density.  In paper II we derive additional constraints 
on the coronal parameters from the Gudel-Benz relation, $L_{\rm radio}/L_{\rm X-ray}\sim 10^{-5}$, which
RQ AGN follow.

\end{abstract}

\section{Introduction}\label{sec:intro}

What is the origin of the radio emission in Radio Quiet (RQ) AGN? In Radio Loud (RL) 
AGN the radio emission is often spatially resolved with a jet like structure, which on milli-arcsec (hereafter mas) scales often shows super luminal motion indicative of bulk relativistic motion. In radio quiet
AGN \citep{kellermann_89,miller_93}, a significant fraction of the radio emission is often unresolved
\citep{kellermann_94,kukula_98,leipski_2006,zuther_2012}, even on pc scale \citep{blundell_98,caccianiga_2001,ulvestad_05,doi_2013,panessa_2013}. 
The observed spectral slopes are often flat, or even inverted \citep{barvainis_96,kukula_98,barvainis_2005,ulvestad_05,Behar_2015}, which indicates the emission is 
not just optically thin synchrotron emission, but 
rather includes some contribution from a compact optically thick source, which can produce a flat or inverted
spectrum.

Some of the radio emission in RQ AGN may be produced by starburst activity in the host galaxy 
\citep{padovani_2011,condon_2013}, however
 the bulk of the radio emission likely originates in the AGN activity 
\citep{white_2015,zakamska_16}. The radio emission may be spatially extended
if it originates in an AGN driven wind which shocks the host galaxy gas 
\citep{gallimore_2006,jiang_2010,ishibashi_2011,zakamska_14,nims_2015}.
The radio may 
also originate from 
a scaled down compact jet 
emission, where the typical jet power is $10^3$ smaller than in RL AGN 
\citep{falke_95, wilson_1995}. 
Observations 
of nearby RQ Seyfert galaxies resolve the emission \citep{giroletti_2009,doi_2013}, and show sub relativistic motion on pc scale \citep{middelberg_2004,ulvestad_2005b}. 
Indications for a compact pc scale source size
are also given by variability \citep{wrobel_2000,anderson_2005,barvainis_2005,mundell_2009}.

Laor \& Behar (2008, hereafter LB08), noted that RQ AGN follow the Gudel-Benz relation,
where $L_{\rm R}/L_{\rm X}=10^{-5}$, which coronally active stars follow. Since the X-ray
emission in RQ AGN may originate in a corona above the accretion disc, it is 
natural to assume that the radio emission in RQ AGN may also originate in the corona, as it does in
coronally active stars.
Although the physical mechanism which leads to the Gudel-Benz relation is not 
understood yet, and the luminosities associated with AGN and coronally active stars deviate
by about 13 orders of magnitude ($10^{31}$ vs. $10^{44}~$erg~s$^{-1}$), the local
coronal conditions may be similar (effective temperature, 
large rotational shear, local densities, see discussion in Gudel 2002). So, it is plausible to expect that similar
mechanisms operate in both systems. The maximal possible synchrotron surface emissivity
(i.e. brightness temperature) implies that the minimal size of the GHz emission region
is on parsec scales (e.g. LB08, eq.22 there), as also suggested by the observed variability
\citep{barvainis_2005}.
Relativistic electrons may reach this radius through the equivalent of Coronal
Mass Ejections (CME) observed in coronally active stars, i.e. magnetized plasma ejected from the innermost
accretion disc. Alternatively, the electrons may reside in a corona which forms a thin layer above an accretion disc, which extends out to pc scale. Although the disc on pc scale must be cold, it may still have a surface corona,
as also seen in cool M stars which are sometimes coronally active \citep{gudel_2002}. 

Synchrotron emission on pc scale, or smaller, may also be produced by a scaled down, low power, jet emission. 
The difference between a CME and jetted emission is more on the descriptive level, where
CME is likely to be poorly collimated and form a sub relativistic outflow, while the term jet is used for a collimated relativistic outflow. Some relativistic jet models in RL AGN invoke a disc corona as the base of the
jet \citep{merloni_2002}, and in RQ AGN the subsequent acceleration may be missing.

Interestingly, high resolution pc scale imaging of NGC~1068 reveal radio 
emission aligned with the central obscuring torus \citep{gallimore_97,gallimore_2004}, which provides 
support for the coronal disc emission scenario for the pc scale radio emission. However, such a pc
scale radio emitting corona is clearly not the dominant source of the X-ray emission in unobscured AGN, 
given the
observed X-ray variability on time scales of days or shorter. The observed X-ray emission must come
from the innermost accreting region, possibly from an X-ray emitting corona above the innermost disc at a
few gravitational radii. Given the maximal intensity of synchrotron emission, the radio emitting corona should be $\sim 100$ times larger than the X-ray 
emitting corona. However, since the minimal
size of the synchrotron emission region scales as $\nu^{-1}$ (LB08), the radio emission
at a few hundred GHz can come directly from the X-ray corona \citep{inoue_2014}.

VLBI observations \citep{blundell_98,caccianiga_2001,ulvestad_05} yield lower limits 
on the brightness temperature of $T_B>10^8-10^9$~K. Although the synchrotron emission is commonly 
assumed to originate in relativistic electrons with a power-law energy distribution, 
this limit on $T_B$ is also consistent with synchrotron emission of thermal electrons
in the radio corona, for a corona temperature of $T\sim 5\times 10^9$~K, as measured for the
X-ray corona \citep{fabian_2015}.

The purpose of this paper is to calculate the possible range of the radio emission 
signatures of coronal synchrotron models. These predictions can be used to explore the validity of 
the coronal emission scenario, and to possibly probe the physical parameters of the corona.  
In section 2 we describe the theoretical background, in section 3 the
numerical solution scheme, in section 4 we provide useful analytic solution, and in section 
5 we present the numerical solution. The discussion is given in section 6, and the conclusions 
are summarized in section 7. 
In paper II (Raginski \& Laor, in preparation) we calculate the implied X-ray emission of the
coronal models used here, and discuss the additional constraints provided by the X-ray observations.

	\section{THE THEORETICAL BACKGROUND}\label{section_theory}
Electrons accelerate and radiate radio emission either when they pass near charged
particle, or when they propagate in a magnetic field. In the earlier case, the free-free
spectral slope $\eta ={\rm dlog}F_{\nu}/{\rm dlog}\nu $ is $\sim -0.1$ from the radio
to $h\nu\sim kT$. The thermal free-free emission is not viable in objects with a power-law 
emission with a steeper
of a flatter $\eta$. Free-free emission of $T>10^7$~K gas is also excluded as
it over predicts the observed X-ray luminosity (e.g. LB08, section 3.6.1 there).  
Free-free is a viable mechanism if the gas is cooler than $10^7$K, but this
is generally ruled out by the VLBI limits of $T_B>10^8-10^9$~K on the core emission.
Thus, electrons in a magnetic field is the only generally viable mechanism.
 Below we briefly review the emission
of relativistic electrons with a power-law (PL) energy distribution (synchrotron emission),
thermal electrons (cyclotron emission), or mildly relativistic thermal electrons 
(cyclo-synchrotron emission, hereafter thermal synchrotron emission). We provide expressions for the emission, absorption,
and radiation transfer used to derive the observed emission.

	\subsection{The synchrotron emission and absorption for thermal and PL electrons}  \label{sec:p_theory}
An electron in a magnetic field $B$ gyrates around the field lines at a frequency 
	\begin{equation}
	\omega_{\rm 0}=\frac{eB}{\gamma m_ec}\equiv\frac{\omega_{\rm B}}{\gamma } , \label{eq:omega_0}
	\end{equation}
where $e$ is the electron charge, $m_e$ is the mass of the electron, $\gamma = \frac{1}{\sqrt{1-\beta^2}}$ is the Lorentz factor, and $\beta=\frac{v}{c} $, where $v$ is the velocity of the electron. Non relativistic electrons ($\beta\ll 1$) radiate cyclotron emission
at $\omega_{\rm B}$. For mildly relativistic electrons ($ \gamma \approx 1  $), additional harmonics appear and the emitted spectrum becomes polychromatic with a few additional lines at higher harmonics. As the electrons become more relativistic ($ \gamma \gg 1 $ ), higher harmonics appear. The width of each line becomes wider with increasing harmonic number, and finally for high enough value of  $\gamma $ and harmonic number $s$, adjacent harmonics merge and a continuous spectrum is generated. 
	
	The resonant frequencies occur at \citep{Zheleznyakov}.
	\begin{equation}
	\omega_{\rm s}=\frac{s\omega_{\rm 0}}{(1- \betam \cos\alpha)}, \label{eq:omega_s}
	\end{equation}
	where $\betam$ is the projection of $\beta $ on the direction of the magnetic field,  and $\alpha$ is the angle between the magnetic field and the observer. The pitch angle $\theta_{\rm p}$ is the angle between the direction of motion of the electron and the magnetic field, which gives 
	\begin{equation}
	\begin{aligned}
		\betam=\beta \cos(\theta_{\rm p})\\
		\betan=\beta \sin(\theta_{\rm p}) \ . \label{(eq:betas)}
	\end{aligned}
	\end{equation}
Equation \ref{eq:omega_s} is correct when the refraction index of the medium is unity, which is a good approximation for AGN coronae, as the frequencies of interest are well above the gyration and Langmuir frequencies ($\sim 300 \rm MHz$ for typical condition assumed). We have verified this numerically by solving for the emission spectrum including the corona refractive index.

We used Zheleznyakov (1970, eqs. 26 \& 35 there) for the emitted energy per harmonic number per solid angle  (erg~sec$^{-1}$~strd$^{-1}$). The specific emission power (erg~sec$^{-1}$~strd$^{-1}$~frequency$^{-1}$) for a unity 
refraction index is \citep{mahadevan_96} 
	\begin{multline}
	\frac{{\rm d}\epsilon}{dt{\rm d}\Omega {\rm d}\omega}=\sum_{s}^{\infty}\frac{e^2\omega^2[\betan^2 J_{\rm s}'^2(\xi)+(\frac{\cos\alpha-\betam}{\sin\alpha})^2 J_{\rm s}^2(\xi)]}{2\pi c}\\
	\times\delta[s\omega_{\rm 0}-\omega(1-\betam \cos\alpha)] , \label{psingle}
	\end{multline} 
	where $\omega$ is the continuum angular frequency, $J_{\rm s}$ is a Bessel function of order $s$, $J'_s$ is the first derivative of the Bessel function of order $s$, and  $\xi$ is 
	\begin{equation}
	\xi=\frac{\omega\betan \sin\alpha}{\omega_{\rm 0}}. \label{xi_ex}
	\end{equation}
	Equation \ref{psingle} provides the emission spectrum of a single electron, with kinetic energy of $(\gamma-1)mc^2$, propagating at a pitch angle of $\theta_{\rm p}$ to the magnetic field. In order to get an expression for the emission per unit frequency per unit volume, 
i.e. the emission coefficient $P(\omega)$, we need to multiply eq.\ref{psingle}
by $n_{\gamma\theta_{\rm p}}$, the number of electrons with velocities in the range of  
$[\gamma,\gamma+{\rm d}\gamma]$ and a pitch angle in the range of $[\theta_{\rm p},\theta_{\rm p}+{\rm d}\theta_{\rm p}]$,
and integrate over $\gamma$. We then integrate over $\theta_{\rm p}$, and $\alpha$, by assuming
a uniform distribution in both angles, i.e. a random distribution of both the electron velocities 
and the magnetic field line directions in a given volume, 

	\begin{multline}
	P(\omega)=\int\limits_{\alpha}\int\limits_{\gamma}\int\limits_{\theta_{\rm p}}p_{\rm single}(\gamma,\omega,\theta_{\rm p})n_{\gamma\theta_{\rm p}} {\rm d}\gamma {\rm d}\theta_{\rm p}\\
	\times 2\pi \sin\alpha {\rm d}\alpha    \label{eq:p_coeff}
	\end{multline}
in units of erg~sec$^{-1}$~cm$^{-3}$~Hz$^{-1}$, 	
where $p_{\rm single}(\gamma,\omega,\theta_{\rm p})=\frac{{\rm d}\epsilon}{dt{\rm d}\Omega {\rm d}\omega}$ is the emission of a single electron. In the next sections we will use
	\begin{equation}
	p_{\rm single}(\gamma,\nu,\theta_{\rm p})=2\pi p_{\rm single}(\gamma,\omega,\theta_{\rm p}) \label{eq:psinge_nu}
	\end{equation} 
	and
	\begin{equation}
	P(\nu)=2\pi P(\omega) . \label{eq:pnu}
	\end{equation}

	An expression for the absorption coefficient is given by  Rybicki \& Lightman 
(2004, hereafter RL04, eq. 6.50 there). We modify the equation to include the dependence on pitch angle $\theta_{\rm p}$,
which gives  
\begin{equation}
\alpha_{\nu}=-\int {\rm d}\theta_{p}\frac{c^{2}}{8\pi\nu^{2}}\int dE \times p_{\rm single}(\gamma,\nu,\theta_{\rm p})E^{2}\frac{\partial}{\partial E}\left(\frac{n\left(E,\theta_{\rm p}\right)}{E^{2}}\right)\ ,\label{eq:RL_abs1}
\end{equation}
where E is the kinetic  energy of the electron, and $n(E,\theta_{\rm p}) $ is the density of electrons per unit energy per pitch angle.

To calculate the radiative transfer we divide the medium into $m$ 
unit volumes (see below the assumed geometry)
of uniform emission and absorption coefficients, say $P_{\rm m}(\nu)$ and $\alpha_{\nu m}$.
If a ray with an intensity  $I_{m-1}$ enters this unit volume, and travels a distance of $d_m$ within this volume,
then the intensity of the exiting ray is 
\begin{equation}
I_m(\nu)=I_{m-1}(\nu)e^{-\alpha_{\nu m}d_{\rm m}}+\frac{P_{\rm m}(\nu)}{4\pi\alpha_{\nu m}}\left(1-e^{-\alpha_{\nu m}d_{\rm m}}\right)\label{eq:I_theory_rec}
\end{equation}
where the first term represents the absorption of the incident intensity, and the second one is the contribution of the emission inside the volume. 
	
\subsection{The geometry} \label{sec:geo_assump}
	\subsubsection{Disc Configuration} 
We assume an optically thick accretion disc which extends from the innermost stable orbit $R_{\rm in}$ at
	\begin{equation}
	R_{\rm 0}\equiv 3R_{\rm S}\ , \label{eq:R0}
	\end{equation}
	where $R_{\rm S}$ is the Schwarzschild radius,  
	\begin{equation}
	R_{\rm S}=\frac{2GM_{\rm BH}}{c^2}\ , \label{eq:Sch_rad}
	\end{equation} 
\citep{shakura_sunyaev_73}. 
	We use $M_{\rm BH}=10^8M_{\sun}$ for the black hole (BH) mass. 
	For the outer boundary of the disc we use $R_{\rm out}=3$~pc, which corresponds to  
$3\times10^5R_{\rm S}$. 
	The observed UV spectral energy distribution suggests a maximal temperature $\sim 5 \times 10^4K$ \citep{laordavis_2014}, which commonly implies a thin disc with an inner radius $R_{\rm in}>R_{\rm 0}$. Therefore, in section \ref{sec:radio_parameters} we study the effect of a larger $R_{\rm in}$ on the radio emission
spectrum. We assume a blanket - like corona that covers the accretion disc, with a uniform thickness of $H=10R_{\rm S}$.

	The thermal electrons density is assumed to scale with radius as
	\begin{equation}
	N(R)=N_0\times (\frac{R}{R_{\rm 0}})^{-q} \label{eq:Nradial}\ .
	\end{equation}
	We generally assume $N_0= 10^9$~cm$^{-3}$, in order to obtain an optical depth for electron scattering of $\tau_{es} \approx 0.1-0.5$ at $R\approx R_{\rm 0}$, as suggested by the X-ray emission (see paper II). In section~5.3 we explore
the dependence of the radio spectrum on the value of $q$.
	
	In addition to the thermal electrons, we assume that the corona also has non thermal electrons
with a PL energy distribution. The PL distribution may be generated during reconnection events, which may also set the temperature of the thermal component.
	The energy distribution of the PL and thermal populations are
		 \begin{equation}
		 \begin{array}{c}
		 n_{\rm pl}(\gamma){\rm d}\gamma=C_{\gamma}\gamma^{-\delta}{\rm d}\gamma      \\
		 \\
		 n_{\rm th}(\gamma) {\rm d}\gamma=\frac{N\gamma^2\beta}{\Theta K_2(\Theta)} e^{-\frac{\gamma}{\Theta}}  
\label{eq:fdist_gamma} ,
		 \end{array}
		 \end{equation}
		 where $C_{\gamma}$ is a normalization constant, $\delta$ is the power index of the PL distribution (typically $\delta=2$ or 3, see section \ref{sec:radio_parameters_pl}), and $\Theta$ is the normalized temperature 
$\Theta \equiv \frac{kT}{mc^2}$, and $K_2$ is the modified Bessel function of the second kind. The expression for
$n_{\rm th}(\gamma)$ is the Maxwell - Juttner distribution, which is the relativistic form of Maxwell Boltzmann. 

		 The energy density of the two populations is assumed to be in equipartition, i.e.	
	\begin{equation}
	\int\limits_{\gamma_{\rm min}}^{\gamma_{\rm max}}C_{\gamma} (\gamma-1)mc^2\gamma^{-\delta}{\rm d}\gamma=\frac{3}{2}NkT \label{eq:pl_eq} \ ,
	\end{equation}
which is used to derive the value of $C_{\gamma}$.	 
	
	The electron temperature in the corona is generally assumed to be $T=5\times 10^9K$, at all radii.
This temperature is naturally expected for a two temperature corona \citep{di_matteo_97}, for
a corona heated by magnetic reconnection \citep{dimatteo_98}, and also for a pair plasma corona
\citep{svensonn_84,Lightmann_zdiarski_87,Haardt_91,Haardt_93}. In addition, recent 
{\em NuSTAR} hard X-ray spectroscopy of AGN \citep{fabian_2015} indeed provide direct evidence for 
a corona with $kT\sim 0.1m_ec^2$, i.e. $T\sim 5\times 10^9$K. However, we also explore below the effect
of using a lower $T$.
 	
	We assume that the local value of the magnetic field within the corona is in equipartition with the thermal electrons,
i.e.
	\begin{equation}
	\frac{B^2(R)}{8\pi}=\frac{3}{2}N(R)kT\ . \label{eq:MF_eq}
	\end{equation}
	
In the numerical calculations we also explore deviations from equipartition, in addition to exploring the effects
of different values for $N_0$, $q$, $\delta$, and $R_0$.

\subsubsection{Spherical Configuration}
For this configuration, we assume a spherical symmetry around the BH. The corona extends for a range of radii $ R_{\rm 0} <R <R_{\rm out}$. The temperature is fixed at $T=5\times 10^9$K, and $N$ is given by eq.\ref{eq:Nradial}. The synchrotron 
emission of both thermal and PL distributions are calculated as described in section \ref{sec:numerical_scheme}.

\section{The numerical solution scheme}
\label{sec:numerical_scheme}

For the thermally distributed electrons, $n_{\rm th}(\gamma){\rm d}\gamma $ is set up using 
eq.\ref{eq:fdist_gamma} for $1.01<\gamma<10$, using 180 bins uniformly spaced logarithmically. 
For the PL electrons, $n_{\rm pl}(\gamma){\rm d}\gamma $ is derived for $1.1<\gamma<3000$
with 300 bins uniformly spaced logarithmically. The upper value of $\gamma$ is selected to ensure
that the peak emission of individual electrons, $\nu=4.1\gamma^2 B$~MHz (RL04, eq.6.17c there), 
extends to $\nu>100$ GHz for the values of $B$ used here.
The pitch angle $\theta_{\rm p}$ is spanned linearly in 14 bins between 0 and $\pi$. 

The number of electrons per unit pitch angle per unit energy are
\begin{equation}
n_{\rm th}(\gamma,\theta_{\rm p}) {\rm d}\gamma {\rm d}\theta_{\rm p}=n_{\rm th}(\gamma){\rm d}\gamma \sin\theta_{\rm p} {\rm d}\theta_{\rm p} /2 \label{eq:n_thermal}\ ,
\end{equation}
and
\begin{equation}
n_{\rm pl}(\gamma,\theta_{\rm p}) {\rm d}\gamma {\rm d}\theta_{\rm p}=n_{\rm pl}(\gamma){\rm d}\gamma \sin\theta_{\rm p} {\rm d}\theta_{\rm p} /2 \label{eq:n_pl}\ .
\end{equation}

\subsection{The emission and absorption coefficients} 
The analytical expression for the spectrum emitted by a single electron (eq.\ref{psingle}) 
includes a delta function. We use the following approximation for the delta function \citep{mahadevan_96}
\begin{equation}
\delta[s\omega_{\rm 0}-\omega(1-\betam \cos\alpha)]=\frac{f(\chi)}{\omega_{\rm B}(1-\betam \cos\alpha)},
\label{eq:delta_approx}
\end{equation}
where $\omega_{\rm B}$ is the Larmour frequency (eq.\ref{eq:omega_0}), and $f(\chi)$ is
\begin{equation}
f(\chi)=\frac{15}{16\Delta\chi}[1-(\frac{2}{\Delta\chi^2})(\chi - \chi_{\rm s})^2+(\frac{1}{\Delta\chi^4})(\chi - \chi_{\rm s})^4,] \label{eq:del_approx_fx}
\end{equation}
where $\chi$ and $\chi_{\rm s}$ are 
\begin{equation}
\begin{array}{c}
\chi=\frac{\omega}{\omega_{\rm B}}\\
\chi_{\rm s}=\frac{\omega_{\rm s}}{\omega_{\rm B}}\ .
\end{array}
\end{equation}
 For the harmonic line width we use $\Delta\chi=0.05\chi_{\rm s}$ \citep{mahadevan_96}.
Combining eq.\ref{eq:delta_approx} and eq.\ref{psingle} we get the emission per single electron
\begin{multline}
p_{\rm single}(\gamma,\omega,\theta_{\rm p})=\sum_{s}^{\infty}\frac{e^2\omega^2[\betan^2 J_{\rm s}'^2(\xi)+(\frac{\cos\alpha-\betam}{\sin\alpha})^2 J_{\rm s}^2(\xi)]}{2\pi c \omega_{\rm B}(1-\betam \cos\alpha)}\\
\times f(\frac{\omega}{\omega_{\rm B}}). \label{eq:psingle_numeric}
\end{multline}
In order to calculate the emission coefficient per frequency, we need to numerically integrate over the relevant electron energy distribution, pitch angle, and observer angle $\alpha$ (eq.\ref{eq:p_coeff}). 
The absorption coefficient is calculated using eq.\ref{eq:RL_abs1}. 

Figure 1 compares the numerical solutions for the emission and absorption coefficients described above, 
with the analytical expression given by RL04 (eq.6.36 there) for the PL 
energy distribution. The analytical and the numerical solutions fit well for $\nu>10^9$~GHz, the region
where the emission and absorption are produced by $\gamma\gg 1$ electrons, where the analytical approximation
is valid. The numerical solution deviates from the analytic one at low $\nu$ since the analytic calculation applies 
only at $\gamma>>1$. We verified the validity of our calculation for the thermal distribution
by comparing our results with the results of Wardzinski \& Zdziarski (2000, figure 5 there), which our results 
overlap.
		
\begin{figure}
			\includegraphics[width=84mm]{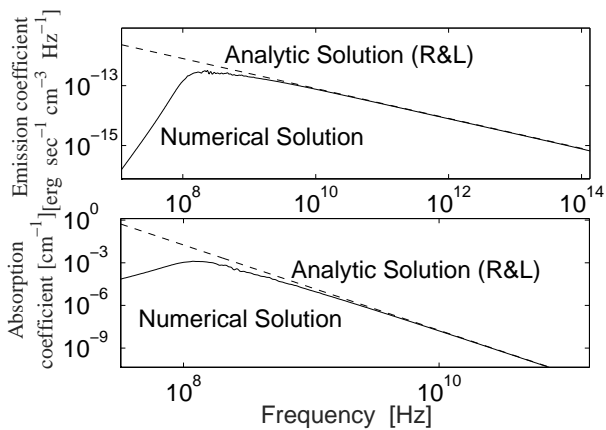}
			\caption{A comparison between numerical and analytical results for the emission and absorption coefficients for PL electrons. The model assumes $B$ which is in equipartition with a thermal component, where  
$N=10^9$~cm$^{-3}$ and $T=10^9$~K (eq.\ref{eq:MF_eq}), and a PL electron distribution with
$\delta =2$, which is also in equipartition (eq.\ref{eq:pl_eq}). The numerical results overlap well
the analytic results (RL04) at $\nu>10^9$~GHz, the region dominated by the emission
of $\gamma\gg 1$ electrons, where the analytic approximation applies.}
			\label{fig:p_and_alpha_RL}
\end{figure}		
	
\subsection {The integrated Radio emission from a disc corona}

\begin{figure}

	\includegraphics[width=84mm]{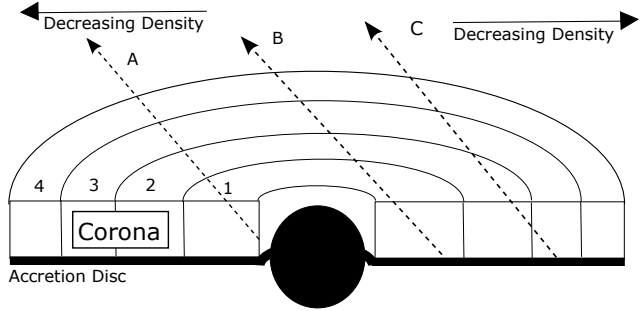}
	\caption{An illustration of the geometry used for the disc corona. The corona is divided into rings, and three rays (A, B and C), are emitted from the surface to the same direction (and thus seen by the same observer). Ray A is emitted from the first ring, and traced back to its origin at the inner wall of the corona. Ray B originates and is emitted from ring 1. Ray C starts in the third ring, and emitted from the second.}
	\label{fig:illus_disc_cross}
\end{figure}

Given the emission and absorption coefficients presented above, 
we  now calculate the total emission of the disc corona. Figure 2 presents the geometry assumed in order to integrate the emission along various lines of sight.
We start with a ray emitted at a given angle $\theta_{\rm e}$ and azimuth
angle $\phi_{\rm a}$, from a given position on the upper surface of the disc corona. The contribution to the intensity is integrated along the ray 
inside the corona until it reaches the face of the underlying optically thick accretion disc, or the sidewalls of the coronal disc (see Fig.\ref{fig:illus_disc_cross}). The integration stops inwards of the innermost part of the disc, $R<R_{\rm 0}$, which represents 
light trapping by the BH. The path of the ray is divided into segments, where every segment is a track of the ray inside a specific coronal ring. We use eq.\ref{eq:I_theory_rec} to calculate the change in intensity following the passage of each segment.
The radii of the coronal rings are logarithmically distributed between $R_{\rm 0}$ and $R_{\rm out}$, according to the values presented in section \ref{sec:geo_assump}. In case of emission from PL electrons, the corona is divided into 25 rings, where the outermost rings is at $R_{\rm out}=3\times10^{5}R_{\rm S}$. For thermal electrons, the corona is extended up to $R_{\rm out}=5\times10^4R_{\rm S}$, and divided into 40 rings. 

The intensity of a specific ray then is multiplied by a geometrical factor of $\cos \theta_{\rm e}$ that stands for the inclination of the emitting area element. The next step is to integrate the emission
from a given ring. This integration is equivalent to a sum of all rays emitted azimuthally from the same point of "coronal face", i.e. an integral over the azimuthal angle $\phi_{\rm a} \in [0,2\pi]$.
\begin{equation}
F_{\rm i}(\nu)= \sum_{\phi_{\rm a}=0}^{2\pi} I_{\rm ray}(\nu) \cos(\theta_{\rm e}) \Delta \phi_{\rm a} ,\label{eq:F_i_nu}
\end{equation}
where $F_{\rm i}(\nu)$ is the flux emitted at a given angle of inclination, from a ring designated by $i$, and
$I_{\rm ray}$ is the intensity of a specific ray emitted from the corona, calculated according to 
eq.\ref{eq:I_theory_rec}. 

We multiply the flux by $R\Delta R$, to get the ring emission (the factor of $2\pi$ is included above in the
integration over $\phi$). We then integrate on radii $R_{\rm 0}<R<R_{\rm out}$ to obtain the total luminosity of the disc. 

\begin{equation}
L_{\nu}=4\pi  \sum_{R_i=R_{\rm 0}}^{R_{\rm out}} R  F_{\rm i}(\nu) \Delta R_i, \label{eq:Lnu_numeric}
\end{equation}
where $L_{\nu}$ is the total radio luminosity of the disc corona at inclination angle of $\theta_{\rm e}$, and the factor of $4\pi$ is used to calculate the inferred isotropic emission even in a non-isotropic case.  

\subsection{The integrated Radio emission from a spherical corona}

\begin{figure}
 	\includegraphics[width=48mm]{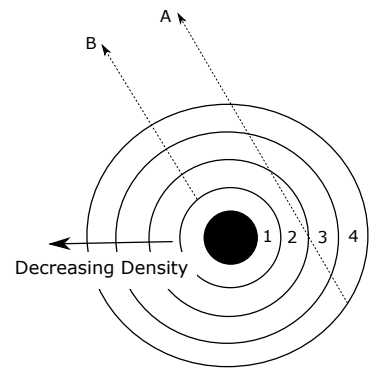}
 	\caption{An illustration of the geometry used for the spherical corona, which is divided into spherical shells. Two rays  are drawn, all of them are emitted towards the same observer. Ray B originates on the innermost shell, as the radiation path cannot pass the BH. }
 	
 	\label{fig:illus_sp_cross}

 \end{figure}

Figure \ref{fig:illus_sp_cross} presents the geometry assumed in the spherical case. The corona is divided into
spherical concentric shells, each with a uniform $B$ and $N$, while the value of $T$ is the same at all shells.
The intensity from a given unit area on the face of the outer shell, is calculated by shooting rays back in all directions. Given the azimuthal symmetry of the radiation transfer, we need to sample only 
$\theta_{\rm e} \in [0, \pi/2]$.
Each ray is linearly traced within the corona, and its intensity is calculated according to eq.\ref{eq:I_theory_rec} for all segments along its path, as is done for the case of disc corona.
 Knowing the intensity per frequency for each ray, we multiply the result by a Jacobian factor of $ \sin\theta_{\rm e} \cos\theta_{\rm e}$ and by a factor of $2\pi $ for integration on azimuthal direction.
All the rays are summed, and their intensities are numerically integrated on angles to obtain the flux.
\begin{equation}
F_{\nu}=\sum_{\theta_{\rm e}=0}^{\pi/2} 2\pi I_{\rm ray}(\nu) \cos(\theta_{\rm e})  \sin\theta_{\rm e}\Delta \theta_{\rm e}
\end{equation}
 The derived flux per unit frequency is then multiplied by a factor of $4\pi R_{\rm out}^2$ to obtain the luminosity. 
 \begin{equation}
L_{\nu}=4\pi R^2_{\rm out} F_{\nu}
 \end{equation}
 
A major difference between the disc and the sphere is that in the disc case we see directly
the emitting surface area at each ring, so the integrated emission is to a good approximation just a 
simple superposition of the emission of the individual rings. In contrast to the spherical case, the radiation
passes from the emission radius to the emitting surface. So, we 
see down to the effective “photosphere”, i.e the radius where the optical depth reaches unity, 
at a given frequency. As a result, the contribution to the emission at a given $\nu$ comes from
a larger range in $R$ in the disc case, compared to the sphere case (see section 5.1).

\section {An approximate analytical solution for the radio spectral slope}
\label{sec:app_solution_spectral_slope}

The spectral slope, $\eta\equiv {\rm d}\log L_{\nu}/{\rm d}\log \nu$, of synchrotron emission from a uniform source of 
PL electrons, is $\eta=-(\delta -1)/2$ for optically thin emission, and $\eta=2.5$ for optically thick emission.
For thermal electrons $\eta=2$ in the optically thick Rayleigh-Jeans regime, with a sharp exponential drop when the
emission becomes optically thin. Below we derive approximate analytical solutions for $\eta$ for a non
uniform synchrotron source, following either the disc or the spherical distribution, and for either thermal
or PL energy distributions. The derived expressions can be used to 
link the observed $\eta$ and the structural parameters of the corona.

The overall spectrum of a stratified corona is a superposition of the spectra of all the rings or shells within the corona. Each ring or shell emits synchrotron radiation with a characteristic spectrum which rises with $\nu$ 
in the optically thick part and drops in the optically thin part. The peak occurs at $\nu_{\rm peak}(R)$ 
which corresponds to the optically thick to thin transition. The transition occurs when the optically thick emission curve (blackbody emission in the thermal electrons case), intersects the optically thin  
emission curve, as derived for the local conditions at a given $R$. Below we derive a general expression for 
$\nu_{\rm peak}$ as a function of the properties of the emitting region, for thermal 
and PL electrons. We then use it to derive $\nu_{\rm peak}(R)$ for the specific corona models we use.

To estimate $\eta$, we assume the emission at a given $\nu$ is dominated by the 
emission from the ring/shell where $\nu=\nu_{\rm peak}$. The implied $\eta$ is then derived by 
taking the ratio of the relative increase in $L_{\nu_{\rm peak}}(R)$, the integrated emission from 
$R_{\rm min}$ to $R$, to the ratio of the relative increases
in $\nu_{\rm peak}$, by the emission from a given ring/shell of a width $\Delta R$ at a distance $R$,
i.e. 
\begin{equation}
\eta = \frac{\Delta \log(L_\nu)}{\Delta \log(\nu)}=\frac{\frac{{\rm dlog}\left(L_{\nu_{\rm peak}}(R)\right)}{dR}\Delta R}{\frac{{\rm dlog}(\nu_{{\rm peak}}(R))}{dR}\Delta R}\label{eq:Slope_theory}
\end{equation}

Below, we derive analytic expressions for $\nu_{\rm peak}(R)$ and $L_{\nu_{\rm peak}}(R)$ 
for the PL and thermal electron distributions for the disc and the 
sphere configurations, and substitute them into the expression above to derive an analytic expression for $\eta$.

\subsection{Analytic derivation of $\nu_{\rm peak}$ for a slab of PL electrons }\label{sec:an_peak_freq_pl}

In this section we assume a slab of thickness $H$, which consists of PL electrons with an energy density set by equipartition with thermal electrons $E=\frac{3}{2}NkT$ . 
The absorption coefficients is (RL04)
\begin{multline}
\alpha_{\nu}=\frac{\sqrt{3}q^{3}}{8\pi m}\left(\frac{3q}{2\pi m^{3}c^{5}}\right)^{\delta/2}C_{E}\left(B\sin\theta_{\rm p}\right)^{\frac{\delta}{2}+1}\Gamma\left(\frac{3\delta+2}{12}\right) \\
\times \Gamma\left(\frac{3\delta+22}{12}\right)\nu^{-\left(\frac{\delta+4}{2}\right)}\label{eq:abs}
\end{multline}
The equipartition magnetic field $B$, and the particle normalization $C_{\rm E}$, are
\begin{equation}
\begin{array}{c}
	B=\sqrt{12\pi NkT}\\
	 \\
	C_{\rm E}=\frac{3NkT}{2\int_{E_{\rm min}}^{E_{\rm max}}E^{1-\delta}dE} \equiv \frac{3NkT}{2K} ,\label{eq:equipartition_nu_peak}
\end{array}
\end{equation}
where $K$ is the integral in the denominator. 

The turnover (peak) frequency satisfies
\begin{equation}
\alpha_{\nu_{\rm peak}}H=\tau\simeq 1\ ,
\end{equation} which gives
\begin{multline}
\nu_{\rm peak}=\left(\frac{16\pi mK \left(\frac{3q}{2\pi m^{3}c^{5}}\right)^{-\delta/2} \left(\sqrt{12\pi k}\sin\theta_{\rm p}\right)^{-\frac{\delta}{2}-1}}{3\sqrt{3}q^{3}k \Gamma\left(\frac{3\delta+2}{12}\right)\Gamma\left(\frac{3\delta+22}{12}\right)}\right)^{-\frac{2}{\delta+4}} \\
T^{\frac{\delta+6}{2\left(\delta+4\right)}} N^{\frac{\delta+6}{2\left(\delta+4\right)}} H^{\frac{2}{\delta+4}}=\label{eq:nupeak_pl_my}
\end{multline}
\[
=A T^{\frac{\delta+6}{2\left(\delta+4\right)}} N^{\frac{\delta+6}{2\left(\delta+4\right)}} H^{\frac{2}{\delta+4}},
\]
where $A$ is a constant at a given $\theta_{\rm p}$. 

For PL electrons with $\delta=2$ we get

\begin{equation}
\nu_{\rm peak} \approx 1.78 \times 10^{-5} T^{2/3} N^{2/3}  H^{1/3}~{\rm Hz} \label{eq:nu_peak_our_pl}
\end{equation}

It is interesting to compare the above expression to an analogous result presented by Gudel (2002, eq. 22 there), for stellar radio emission

\begin{equation}
\nu_{\rm peak-Gudel}=\left[10^{11.77+3.44\delta}\left(\delta-1\right)^{2}n^{2}H^{2}B^{\delta+2}\right]^{\frac{1}{\delta+4}}~{\rm Hz},
\end{equation}
where $n$ is the number density of the PL electrons (generally $n\ll N$).
Using the above expression, and applying equipartition of the thermal electrons, PL electrons, and the magnetic field, we obtain
\begin{equation}
\nu_{\rm peak-Gudel}\propto N^{\frac{\delta+6}{2\left(\delta+4\right)}}T^{\frac{\delta+6}{2\left(\delta+4\right)}}H^{\frac{2}{\delta+4}} \label{eq:nu_peak_gudel}.
\end{equation}

For $\delta =2$, and applying equipartition, one obtains
\begin{equation}
\nu_{\rm peak-Gudel}\approx 7\times 10^{-6}T^{2/3}N^{2/3}H^{1/3}{\rm Hz },
\end{equation}
i.e. the same functional dependence derived above. The coefficient in Gudel is a factor of 2.5
too small, which may reflect the accuracy of the approximate analytic derivations.

\subsection{Analytic derivation of $\nu_{\rm peak}$ for a slab of thermal electrons }

There is no simple analytic expression for the synchrotron absorption coefficient $\alpha_\nu $ for electrons
with a thermal energy distribution. 
Instead, we use the numerical results to obtain a fitting function for $\nu_{\rm peak}$ as a function of $T$, 
$N$ and $H$. The derived fitting function is 
\begin{equation}
\nu_{\rm peak}=3.74\times10^{-9} T^{1.42}N^{0.55}H^{0.09}~{\rm Hz} \label{eq:our_nupeak_thermal}
\end{equation}
for $T=10^8-10^{10}$K, $N=10^8-10^{10}$~cm$^{-3}$, and $H=10^{14}-10^{15}$~cm.

\cite{gudel_2002} derives a similar expression
\begin{equation}
\nu_{\rm peak-Gudel}=1.3\left(\frac{N H}{B}\right)^{0.1}T^{0.7}B
\end{equation}
which gives for an equipartition $B$
\begin{equation}
\nu_{\rm peak-Gudel}=4.8\times10^{-7}T^{1.15}N^{0.55}H^{0.1}\label{eq:Gud_nupeak}
\end{equation}
Note the difference between the power indices of $T$ in eq.\ref{eq:our_nupeak_thermal} (1.42)
and eq.\ref{eq:Gud_nupeak} (1.15). The calculation in \cite{gudel_2002} is performed only for relatively low harmonic numbers: $10<s<100$, which are relevant for the stellar corona, where $T<10^8$K.  Imposing the same limitations on the harmonic numbers in our simulation, we derive $\nu_{\rm peak}\propto T^{1.1}$, 
in reasonable agreement with the power of 1.15 in \cite{gudel_2002}.  This difference stresses the need
to retain high harmonic numbers in the calculations for mildly relativistic thermal electrons. 

 Additional comparison was performed with the results of an analogous calculation of the peak frequency performed 
by Wardzinski \& Zdziarski (2000, eq.18 there). Applying equipartition to their expression gives 
$\nu_{\rm peak}~\sim 5.2\times 10^{-8}T^{1.45}N^{0.5}H^{0.05}$. The small deviation in the power indices from
our expression (eq.\ref{eq:our_nupeak_thermal})
reflects the uncertainty in the fits to the numerical results. The difference between the coefficients is "compensated" by the small difference in the powers of $T$, leading to similar values of 
$\nu_{\rm peak}$ at the relevant range of temperatures.

\subsection{The spectral slope for emission from a disc}

Below we calculate $\eta$ for synchrotron emission from a disc for both thermal and PL electrons. 
For the sake of simplicity we assume a constant $H$, and for $T$ and $N$ we assume  
\begin{equation}
\begin{array}{c}
T\left(R\right)=T_0\left(\frac{R}{R_{\rm 0}}\right)^{-p},\\
N\left(R\right)=N_0\left(\frac{R}{R_{\rm 0}}\right)^{-q}.\\
\end{array}\label{eq:TnB_on_R}
\end{equation}

\subsubsection{Thermal Electrons}

Substituting the above $N(R)$, $T(R)$, and a constant $H$, into eq.\ref{eq:our_nupeak_thermal} gives
\begin{equation}
\nu_{\rm peak}(R)\propto R^{-1.42p-0.55q} \label{eq:nu_peak_thermal_disc}\ .
\end{equation} 
The spectrum of a ring of thermally distributed electrons is that of Rayleigh-Jeans emission, i.e. rises as $\nu^2$
 in the optically thick range, and is exponentially falling in the optically thin range. The specific luminosity $L_{\nu}$ at frequency $\nu$ is 
\begin{equation}
L_{\nu_{\rm peak}}(R)=\int_{R_{\rm min}}^{R}\pi \mathscr{B}_{\nu_{\rm peak}}(r) 2\pi rdr \label{eq:Lnu_thermal_1}
\end{equation}
where $\mathscr{B}_{\nu_{\rm peak}}(r) $ is the value of the blackbody emission at radius $r$, and at a frequency 
$\nu_{\rm peak}(R)$.
We integrate only on rings inner to $R$, because they are optically thick at a frequency of $\nu_{\rm peak}(R)$ and their contribution is not negligible. The outer rings, at $r>R$, are optically thin at $\nu_{\rm peak}(R)$, and their contribution is very small due to the fast exponential drop of the optically thin thermal synchrotron spectrum. 
Since $h\nu (\sim 10^{-3}{\rm eV})\ll kT(\sim 100{\rm keV})$, the Rayleigh-Jeans approximation applies, 
\begin{equation}
\mathscr{B}_{\nu_{\rm peak}}(r) =\frac{2\nu^2_{\rm peak}(R)kT(r)}{c^2} . \label{eq:Bnu_thermal}
\end{equation}
Substituting eqs. \ref{eq:TnB_on_R}, \ref{eq:nu_peak_thermal_disc} and \ref{eq:Bnu_thermal} into eq.\ref{eq:Lnu_thermal_1}, and setting $R_{\rm 0}/R\ll 1 $, we obtain
\begin{equation}
L_{\nu_{\rm peak}}(R) \propto R^{2-3.84p-1.1q} .\label{eq:L_nu_thermal_disc}
\end{equation}
Substituting eq.\ref{eq:L_nu_thermal_disc} and eq.\ref{eq:nu_peak_thermal_disc} into eq.\ref{eq:Slope_theory}, 
gives
\begin{equation}
\eta_{\rm disc-thermal}=\frac{2-3.84p-1.1q}{-1.42p-0.55q} \label{eq:slope_disc_thermal} .
\end{equation}

\subsubsection{PL electrons}
\label{sec:pl_spectral_slope}
As in the thermal case, for PL electrons we substitute eq.\ref{eq:TnB_on_R} into 
eq.\ref{eq:nupeak_pl_my}, and get
\begin{equation}
\nu_{\rm peak} (R)\propto R^{-\frac{\delta +6}{2(\delta+4)}p-\frac{\delta +6}{2(\delta+4)}q} .\label{eq:nu_peak_pl_disc}
\end{equation}
The luminosity is 
\begin{equation}
L_{\nu_{\rm peak}}(R) = 2\pi\int_{R_{\rm 0}}^{R}\frac{P_{\nu_{\rm peak}}\left(r\right)}{4\pi\alpha_{\nu_{\rm peak}\left(r\right)}}2\pi rdr\label{eq:eq16}\ ,
\end{equation}
where the factor of $\pi$ outside of the integral converts the source function to flux, and the
factor of 2 accounts for the two faces of the disc. This expression holds in areas where the 
corona is optically thick, which holds at $r<R$ for $\nu=\nu_{\rm peak}(R)$.

According to RL04, $P_{\nu_{\rm peak}}$ is given by
\begin{multline}
P_{\nu_{\rm peak}}=\frac{\sqrt{3}q^{3}C_{\gamma}B\sin\theta_{\rm p}}{mc^{2}\left(\delta+1\right)}\Gamma\left(\frac{\delta}{4}+\frac{19}{12}\right)\Gamma\left(\frac{\delta}{4}-\frac{1}{12}\right)  \\ 
\times \left(\frac{2\pi mc\nu_{\rm peak}}{3qB\sin\theta_{\rm p}}\right)^{-\frac{\delta-1}{2}} .\label{eq:P_nu_RL}
\end{multline}
Applying equipartition (eq.\ref{eq:MF_eq}) and the expression for
$C_{\gamma}$ (eq.\ref{eq:pl_eq}) gives
\begin{equation}
P_{\nu_{\rm peak}}(r) \propto n^{\frac{5}{4}+\frac{\delta}{4}}T^{\frac{5}{4}+\frac{\delta}{4}}\nu_{\rm peak}^{-\frac{\delta-1}{2}}(R)\ ,
\end{equation}
Applying the radial dependence (eq.\ref{eq:TnB_on_R}) gives
\begin{equation}
P_{\nu_{\rm peak}}(r) \propto r^{-\left(\frac{5}{4}+\frac{\delta}{4}\right)p-\left(\frac{5}{4}+\frac{\delta}{4}\right)q}\nu_{\rm peak}^{-\frac{\delta-1}{2}}(R) . \label{eq: Pcoeff_R}
\end{equation}
The absorption coefficient (eq.\ref{eq:abs}), with equipartition (eq.\ref{eq:MF_eq}), gives
\begin{equation}
\alpha_{\nu_{\rm peak}} \propto  n^{\frac{\delta}{4}+\frac{3}{2}}T^{\frac{\delta}{4}+\frac{3}{2}}\nu_{\rm peak}^{-\frac{\delta+4}{2}}(R) .
\end{equation}
Applying the radial dependence (eq.\ref{eq:TnB_on_R}) gives
\begin{equation}
\alpha_{\nu_{\rm peak}} \propto  r^{-\left(\frac{\delta}{4}+\frac{3}{2}\right)p-\left(\frac{\delta}{4}+\frac{3}{2}\right)q}\nu_{\rm peak}^{-\frac{\delta+4}{2}}(R) . \label{eq:ebs_coeff_R}
\end{equation}
The luminosity density is then (eq.\ref{eq:eq16})
\begin{multline}
L_{\nu_{\rm peak}}(R)=\nu_{\rm peak}^{5/2}(R)\int_{R_{\rm 0}}^{R}\frac{r^{-\left(\frac{5}{4}+\frac{\delta}{4}\right)p-\left(\frac{5}{4}+\frac{\delta}{4}\right)q}}{r^{-\left(\frac{\delta}{4}+\frac{3}{2}\right)p-\left(\frac{\delta}{4}+\frac{3}{2}\right)q}} rdr\\
=\nu_{\rm peak}^{5/2}(R)\int_{R_{\rm 0}}^{R}r^{\frac{p}{4}+\frac{q}{4}+1}dr \ ,
\end{multline}
which gives
\begin{equation}
L_{\nu_{\rm peak}}(R) \propto \nu_{\rm peak}^{5/2}(R)R^{-\frac{p}{4}-\frac{q}{4}+2}\ .
\end{equation}
for $R\gg R_{\rm 0}$.

Applying the expression for $\nu_{\rm peak}(R)$ (eq.\ref{eq:nu_peak_pl_disc}) gives
\begin{multline}
L_{\nu_{\rm peak}}(R) \propto\left(R^{-\frac{\delta+6}{2\left(\delta+4\right)}p-\frac{\delta+6}{2\left(\delta+4\right)}q}\right)^{5/2}R^{\frac{p}{4}+\frac{q}{4}+2}=\\
=R^{-\frac{5(\delta+6)}{4\left(\delta+4\right)}p+\frac{p}{4}-\frac{5(\delta+6)}{4\left(\delta+4\right)}q+\frac{q}{4}+2} , \label{eq:L_nu_peak_disc_pl}
\end{multline}
or
\begin{equation}
L_{\nu_{\rm peak}}(R)\propto R^{-\frac{17}{12}(p+q)+2} , 
\end{equation}
for the $\delta=2$ case. The derived spectral slope (eq.\ref{eq:Slope_theory}) is
\begin{equation}
\eta_{\rm disc-pl}=\frac{17(p+q)-24}{8(p+q)} \label{eq:slope_disc_pl}\ ,
\end{equation}
while for the $\delta=3$ case we get
\begin{equation}
\eta_{\rm disc-pl-\delta-3}=\frac{19(q+p)-28}{9(q+p)} . \label{eq:slope_disc_pl3}
\end{equation}

Note that the above analytic estimate for $\eta$ for PL electrons is valid only at
$\eta>-\frac{\delta -1}{2}$, since this estimate ignores the contribution of the optically thin region, which
sets this lower limit on $\eta$.

\subsection{The spectral slope for emission from a sphere}

The emission from a sphere is qualitatively different from disc emission. In the case of a disc, the 
emission can be considered as a superposition of rings, each one is directly observed. The emission from
a sphere can be considered as a superposition of spherical shells, but the emission of each shell
propagates through all outer shells, and we effectively see only the emission from a volume set by
the surface and the $\tau\approx 1$ radius, deeper shells do not contribute. The thickness of each shell is set by 
the scale length of $B(R)$, i.e. \citep{gudel_2002}
\begin{equation}
H\left(R\right)=\frac{B\left(R\right)}{\left|\nabla B\left(R\right)\right|}\sqrt{\frac{kT}{mc^{2}}} .
\end{equation}
Using the equipartition for $B(R)$, $N(R)$ and $T(R)$ (eqs. \ref{eq:pl_eq},\ref{eq:MF_eq}),
gives $B\propto R^{-\frac{p+q}{2}}$ and $\nabla B \propto R^{-\frac{p+q}{2}-1}$,
which gives
\begin{equation}
H(R) \propto R^{1-\frac{p}{2}}\label{eq:SR_Sphere}\ ,
\end{equation}
which we use below to estimate the spectral slope.

\subsubsection{Thermal electrons}

Applying the expression for $\nu_{\rm peak}(N,T,H)$ (eq.\ref{eq:our_nupeak_thermal}) for the sphere yields
\begin{equation}
\nu_{\rm peak}(R)\propto R^{0.09-1.465p-0.55q} \label{eq:nu_peak_thermal_sphere}\ .
\end{equation}
To derive $L_{\nu}$ we assume the emission originates only from the shell at $R$, which produces a peak at $\nu$,
which gives
\begin{equation}
L_{\nu_{\rm peak}}\left(R\right)=4\pi R^{2}  \pi \mathscr{B}_{\nu}\ ,
\end{equation} 
where $\mathscr{B}_{\nu}$ is the Planck function of a shell $R$. Thus
\begin{equation}
L_{\nu_{\rm peak}}(R) \approx \frac{8\pi^2 h\nu^2_{\rm peak}\left(R\right)kT\left(R\right)}{c^{2}} R^2\propto R^{2.18-3.93p-1.1q} \label{eq:L_nu_thermal_sphere}\ ,
\end{equation}
which gives
\begin{equation}
\eta_{\rm sphere-thermal}=\frac{2.18-3.93p-1.1q}{0.09-1.465p-0.55q} \label{eq:slope_sphere_thermal}
\end{equation}

\subsubsection{PL electrons}
In this case we apply (eq.\ref{eq:nupeak_pl_my}) for $\nu_{\rm peak}$, which gives using
eq.\ref{eq:TnB_on_R} and eq.\ref{eq:SR_Sphere}
\begin{equation}
\nu_{\rm peak}(R) \propto R^{-\frac{4-p\delta-q\delta -8p-6q}{2(\delta +4)}} \label{eq:nu_peak_pl_sphere}\ .
\end{equation}
The luminosity is given by
\begin{equation*}
L_{\nu_{\rm peak}}(R)=\frac{ P_{\nu_{\rm peak}}\left(R\right)}{\alpha_{\nu_{\rm peak}\left(R\right)}} \pi R^2dR \ ,
\end{equation*}
where $P_{\nu_{\rm peak}}$ is the emission coefficient (eq.\ref{eq: Pcoeff_R}) and $\alpha_{\nu_{\rm peak}} $ is the absorption coefficient (eq.\ref{eq:ebs_coeff_R}), which yields
\begin{equation}
L_{\nu_{\rm peak}}(R) \propto R^{\frac{-2p\delta-2q\delta-18p-13q+26+4\delta}{2\left(\delta+4\right)}}\ .
\end{equation}
The implied slope for $\delta = 2 $ is 
\begin{equation}
\eta_{\rm sphere-pl}=\frac{22p+17q-34}{10p+8q-4} . \label{eq:slope_sphere_pl}
\end{equation}

As noted above, the expressions for $\eta_{\rm sphere-thermal}$ and $\eta_{\rm sphere-pl}$ are valid
only for slopes above the optically thin range, i.e. for $\eta>-(\delta-1)/2$.
  
\section{RESULTS} \label{sec:results}

Below we present the results of the numerical calculations of the radio emission. The calculations are
for the disc and spherical geometries, and for thermal and PL electrons. In all cases we assume
an isothermal corona, and a decreasing $N(R)$, i.e. $p=0$, and $q>1$ (eq.\ref{eq:TnB_on_R}). 
The value of $B(R)$ is derived from the assumption of equipartition with the thermal electrons (eq.\ref{eq:MF_eq}). 
The PL electrons are also assumed to be in equipartition with the thermal electrons 
(eq.\ref{eq:pl_eq}). 

The innermost radius of the corona is assumed to be the last stable orbit for a central BH of a mass of $10^8M_{\sun}$ (eqs.\ref{eq:R0}, \ref{eq:Sch_rad}), unless noted otherwise. The outermost radius is taken to be $5\times 10^4 R_{\rm S}$ for the thermal distribution, and $3\times 10^5 R_{\rm S}$ for the PL electrons, which ensures the spectral break 
due to transition to optically thick emission occurs at $\nu< 1$~GHz (see below). As noted above, 
the coronal thickness in the disc configuration is assumed to be constant at $H=10R_{\rm S}$. 

We concentrate below on models which yield a flat spectral slope, as this is the unique signature of
a compact emission source, such as a disc corona. As shown below, for some parameters the optically
thin emission from the outer parts of the corona, can dominate the emission from the inner parts.
In such a case, the overall spectrum is just that of an optically thin source, which is derived
also in other more extended emission models. 

 \begin{figure*}
 	\includegraphics[width=175mm]{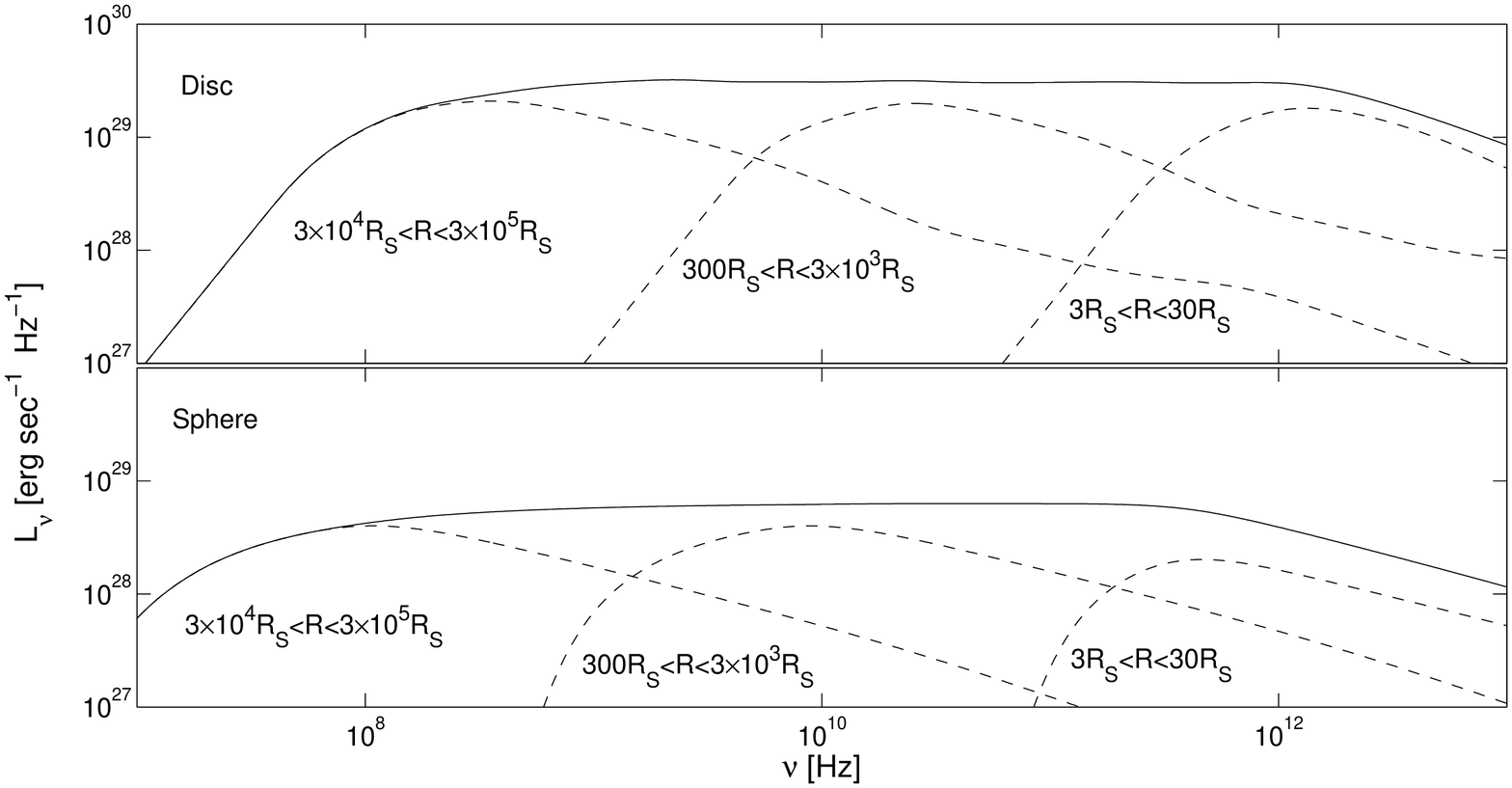}
 	\caption{ The spectrum of PL electrons ($\delta =2$ ) from a corona in a disc (upper panel) 
or a sphere (lower panel). The disc is observed at an inclination angle of $\cos\theta_{\rm e} = 0.5$. 
The solid line in each panel represents the overall emission, and the dashed lines 
the contributions of different rings or shells. The corona is isothermal with $T=10^9K$. The
coronal density drops as a PL with $q=1.4$ for a disc, and $q=2$ for a sphere, selected to produce a 
flat continuum, and $N_0=10^9~{\rm cm}^{-3}$ to set the luminosity scale (see section \ref{sec:results}). 
 The emission below 1~GHz originates mostly at $R>3\times 10^4R_{\rm S}$, i.e. $R>10^{18}$~cm 
for $M_{\rm BH}=10^8M_{\odot}$ used here. At 100~GHz (3~mm) the emission comes mostly from $R\sim 100R_{\rm S}$.
The emission at $\nu>10^{12}$~Hz in RQ AGN is heavily dominated by cold dust emission, but may
be measured with VLBI which detects only high $T_b$ sources. }

 	\label{fig:pl_rings}
 \end{figure*}

  \begin{figure*}
  	\includegraphics[width=175mm]{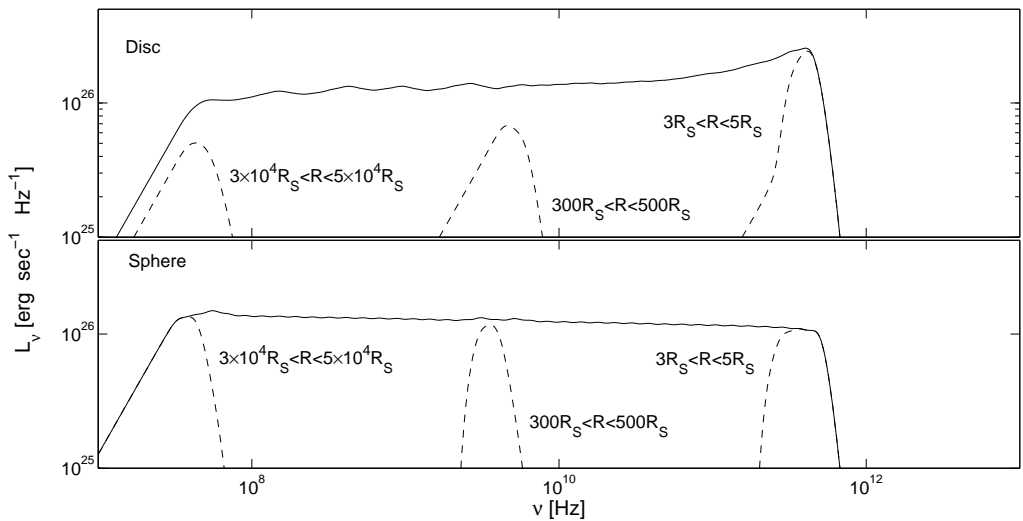}
  	\caption{The synchrotron emission of the thermal electrons at $T=5\times 10^9$K, for the same 
configurations as in Fig.4. The coronal density drops with $q=1.82$ for a disc, and $q=1.98$ 
for a sphere, with $N_0=10^9~{\rm cm}^{-3}$ (see section \ref{sec:results} for further details). 
Although the thermal electrons are in equipartition with the PL electrons, their 
synchrotron luminosity is significantly smaller than that produced by the 
PL electrons. Note also the narrow range of frequencies of the emission from a given
ring or shell, in contrast with the broad distribution of the PL electrons emission.
The wiggles appearing in the spectrum are a numerical artefact caused by the finite resolution
of the radial integration.} 
  	
  	\label{fig:thermal_rings}
  \end{figure*}

\subsection{The radial dependence of the emission} \label{sec:radial_dependence}

Figures \ref{fig:pl_rings} and \ref{fig:thermal_rings} present typical spectra of thermal and PL distributed electrons within a disc or spherical corona. The values of the density distribution $q$ parameters were selected  
to produce a flat spectrum ($L_{\nu} \propto \nu^0$) 
in the intermediate frequency range (1-100~GHz), before reaching the steeply falling optically thin limit at 
higher frequencies, and the steeply rising optically thick emission at lower frequencies. 
The figures also present the contributions to the total emission of different rings or spherical shells. 

The emission of thermal electrons from a given ring or shell (Fig.5), covers a smaller range of frequencies, 
compared to the emission of the PL electrons (Fig.4). Both electron distributions produce similar spectral slopes in the locally optically thick emission at $\nu<100$MHz 
(2 for thermal, 2.5 for PL in a disc), but in the locally optically thin part the thermal emission shows a sharp exponential drop, while
the PL show only the gradual optically thin falloff with a slope of $-1/2$. The sharp cutoff of the 
optically thin thermal synchrotron results from the exponential drop in the electron energy
(eq.14), together with the exponential drop in the emission of a single electron
(eq.6.34b in RL04). 

The slope of the locally optically thick emission in the spherical configuration is steeper than in the disc configuration, as the emission is absorbed by the outer shells.
In contrast with the disc case, where the emission from each ring is directly observed.

  \begin{figure*}
  	\includegraphics[width=175mm]{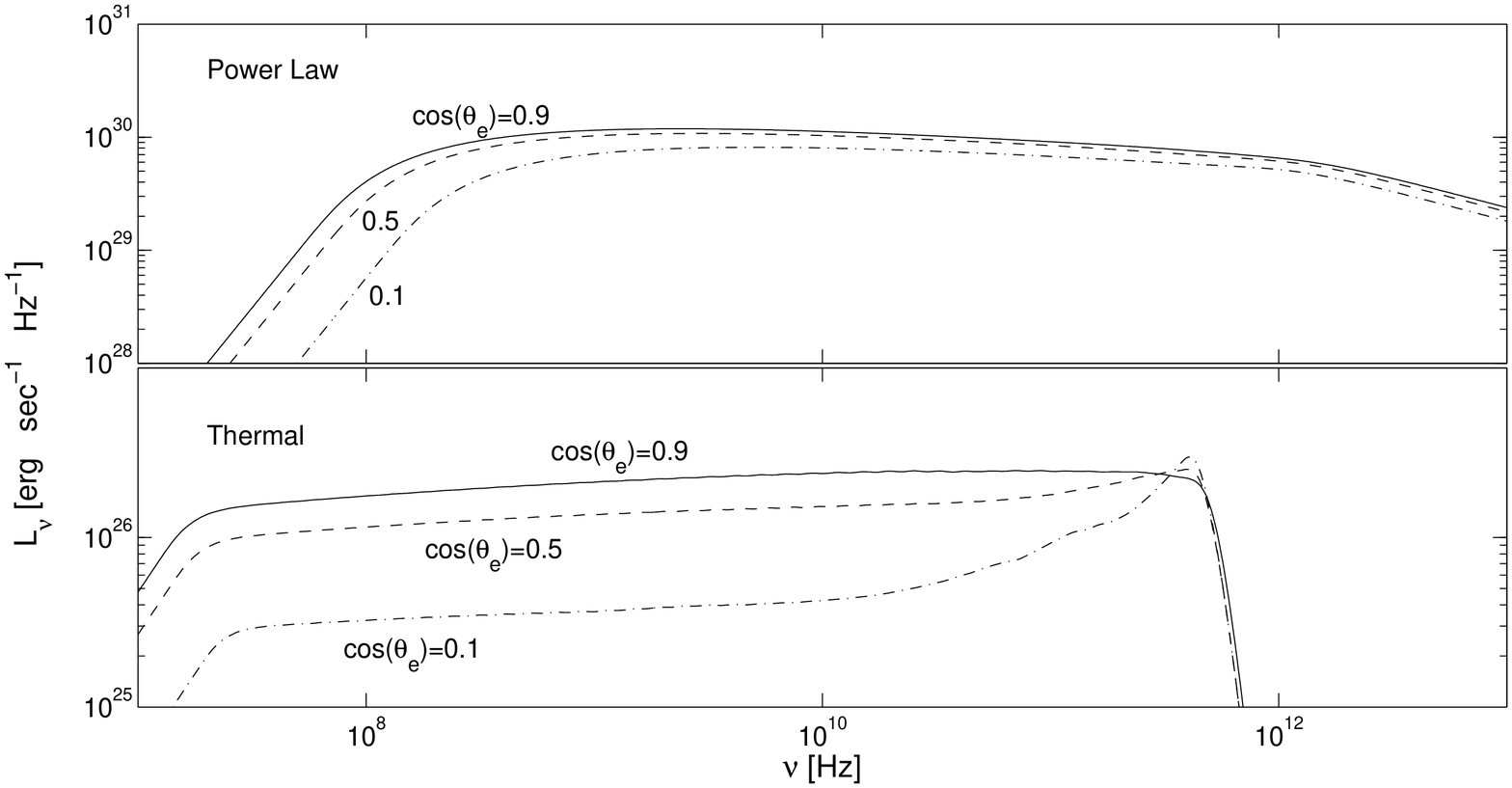}
  	\caption{The inclination dependence of coronal disc emission for thermal and PL electrons. At the low
frequencies the emission is optically thick, which produces a $\cos\theta_{e}$ dependence. At high frequencies
the entire disc becomes optically thin, and the emission becomes isotropic. In the thermal case the exponential drop
when the emission is optically thin leads to negligible emission, and the integrated emission is dominated by the optically
thick regions. The rise in the emission for $\cos\theta_{e}=0.1$ with frequency for thermal electrons is a geometry 
effect (see section \ref{sec:results})}.  	
  	\label{fig:angles}
  \end{figure*}

\subsection{The inclination dependence of the emission} \label{sec:radio_parameters_inclinaion}
  
Figure \ref{fig:angles} presents the observed disc emission for PL and thermal electrons as a function of 
$\cos\theta_{e}$, where $\theta_{e}$ the disc inclination angle. The spectrum can be roughly divided into three regimes, with different inclination dependence. At low enough frequencies, the whole disc is optically thick, and the luminosity scales as $\cos\theta_{e}$. At high enough frequencies the whole disc is optically thin, the emission becomes
isotropic, and the spectra from all inclinations overlap.

At intermediate frequencies, the inner part of the disc is optically thick and an outer part of it is optically thin.
The transition radius decreases with increasing frequency, until the whole disc emission becomes optically thin.
As a result, at intermediate frequencies we get an intermediate inclination dependence. In the thermal case, the exponential drop in the locally optically thin emission, leads to a negligible
contribution from the optically thin isotropically emitting part of the disc. As a result, the optically 
thick inclination dependence remains 
$\cos\theta_{e}$ to almost the highest frequencies.
      
A peculiar behavior occurs for the thermal case at $\nu > 10^{11}$Hz at high inclination (lower panel of Fig.\ref{fig:angles}),
where the luminosity at $\cos\theta_{e}=0.1$ becomes comparable, and even exceeds the emission at $\cos\theta_{e}=0.9$
(for $\nu \approx 5\times 10^{11}$Hz). This results from the adopted geometry, as the highest frequency thermal 
emission comes mostly from the innermost ring, where the height $H=10R_{\rm S}$ becomes larger than the radius. Most of the emitting 
area is now in the walls, and the emission becomes stronger at high inclinations, where the projected area of the walls becomes largest.
Thermal synchrotron emission therefore provides a sensitive measure of the projected surface area of the region where the
observed frequency is produced.

\subsection{The dependence of the emission on the coronal properties} \label{sec:radio_parameters}

\begin{figure*}
  	\includegraphics[width=175mm]{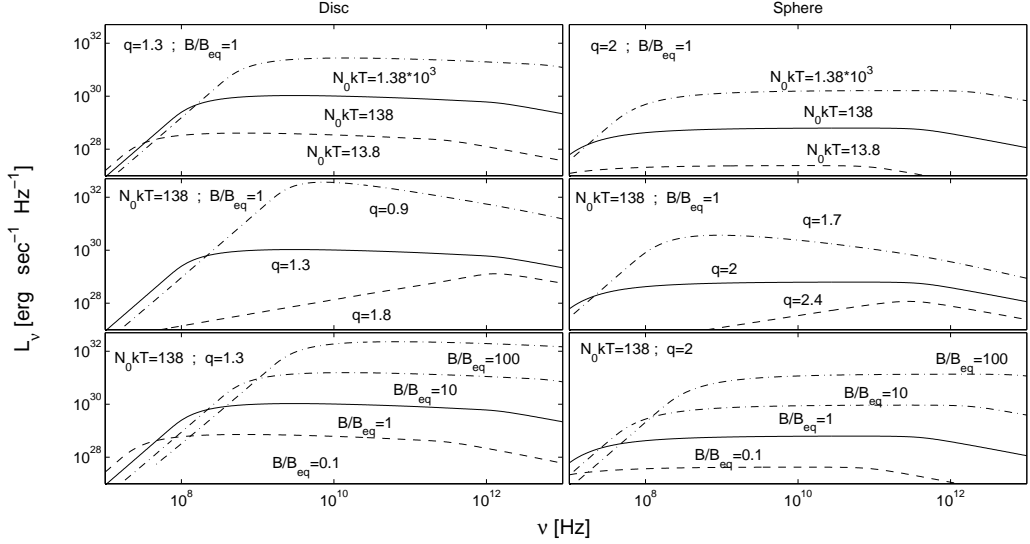}
  	\caption{The spectra of PL electrons in a disc (left panels) and spherical (right panels) coronae. 
All calculation assume $T=10^9K$, $\delta = 2$, $R_{\rm 0}=3R_{\rm S}$, $R_{\rm max}= 3\times 10^5R_{\rm S}$, for $M_{\rm BH}=10^8M_{\sun}$. Each panel
notes the assumed values of $N_0kT$ (in cgs at $R_{\rm 0}$), $q$ of the density distribution, and $B/B_{\rm eq}$. In all cases,
the disc luminosity is higher than the sphere luminosity, for similar parameters.
The upper panels present the dependence on the equipartition energy density, as noted by the value of $N_0kT$. 
The emission remains optically thick (slope of 2.5) to higher frequencies, with increasing $N_0kT$, leading to higher
luminosities in the flat part of the spectrum. The middle panels present the $q$ values required to derive slopes of about 
$1/2$, 0, and $-1/2$ in the flat part of the spectrum. In all cases the $q$ values of the spherical case are steeper
than the disc case. The lower panels present the effect of the deviations of $B$ from equipartition. The luminosity
in the optically thick part decreases with increasing $B$, and rises with $B$ in the flat slope region. Note that some of the
models lead to optically thick emission at $\nu>1$GHz, which is not observed.}
  	
  	\label{fig:pl_parameters}
  \end{figure*}

Figure \ref{fig:pl_parameters} presents the emission of PL electrons from a disc (left panels) or a sphere (right panels).
In all cases we assume a PL index $\delta = 2$, and corona with $T=10^9$K with a constant thickness, which extends 
from $R_{\rm in}=3R_{\rm S}$ to $R_{\rm max}= 3\times 10^5R_{\rm S}$ for $M_{\rm BH}=10^8M_{\sun}$ (which corresponds to $R_{\rm S}=3\times 10^{13}$~cm).
The upper panels present the dependence of the spectrum on the coronal thermal energy density $NkT$, where we assume an equipartition with the
PL electrons energy density, and the magnetic energy density. As $NkT$ increases, i.e. the normalization of $N$ increases,
the luminosity increases, as does the transition frequency from optically thick emission (the 2.5 slope) to a flat continuum.
Note also that the disc luminosity is larger than the sphere luminosity by about an order of magnitude. This results from
the steeper $q=2$ required in the spherical case, compared to $q=1.3$ in the disc case, to derive the same flat continuum slope, using the same $N_0$ in both geometries.

The second row of panels presents the dependence on $q$ (eq.\ref{eq:Nradial}).
The values of $q$, noted near each curve, were selected to produce spectral slopes of $-0.5 , -0.18$ and $ 0.45 $ for a disc, and $-0.5, 0$ and $ 0.45 $ for a sphere, based on the analytical expressions given in eq.\ref{eq:slope_disc_pl} and eq.\ref{eq:slope_sphere_pl}. The slopes derived from the numerical solutions are $-0.42, -0.11, 0.41$ for the disc, and $-0.3 ,0.01, 0.45$ for the sphere, which implies that the analytic estimate is typically accurate to 
better than 0.1. The overall trend is that a larger $q$ yields a flatter slope, as expected as a larger drop
in the density yields a larger drop in the synchrotron emission with radius. Since the outer radius contributes
at lower frequencies, a higher $q$ implies the inner region dominates, leading to a flatter spectral slope.

If $q$ is low enough (0.9 for a disc, 1.7 for a sphere), the emission from the outer disc becomes
dominant enough, that the optically thin tail it produces at higher frequencies dominates the emission 
from the inner disc.
The observed emission is effectively all produced by emission from a single uniform emitter at the 
outer disc, 
rather than by a superposition of emitters at all radii. The spectrum is then a steeply rising 
optically thick spectrum at low enough frequencies, has a peak where the outer disc becomes optically thin,
and shows an optically thin tail at high frequencies, somewhat similar to the
spectra observed in GHz peaked radio loud sources \citep{odea_98,sadler_2015}.

The $q$ values in the spherical case are steeper than in the disc case, for the same spectral
slope. This is expected since the change of the synchrotron emissivity with radius depends on the change 
in the column density with radius. In the disc case $q$ also provides the radial dependence of the column 
density, while in the sphere case the column density scales as $q+1$.

The flat spectrum in the spherical geometry case implies $q=2$ for both PL and thermal electrons. This 
implies that the gas thermal energy density $3NkT$ is a fixed fraction of the radiation energy density
$L/4\pi R^2c$. A bolometric luminosity of $L\simeq 10^{46}$~erg~s$^{-1}$ corresponds to 
$L_{\nu}\simeq 10^{30}$~erg~s$^{-1}$~Hz$^{-1}$ at 10~GHz, which is produced by a corona with 
$NkT=1380$~erg~cm$^{-3}$ at $R=3R_{\rm S}=9\times 10^{13}$~cm for the PL case (Fig.7 upper right panel). 
This implies a fixed energy density ratio of $4\times 10^{-4}$ at all radii.

The third row of panels in Fig.\ref{fig:pl_parameters} presents the effect of deviation from equipartition,
explored by varying $B/B_{\rm eq}$ from 0.1 to 100, while the PL electrons remain in equipartition with the thermal
plasma energy density. As expected, the luminosity in the intermediate range of frequencies 
increases with $B/B_{\rm eq}$. In the optically thick range (low frequencies), we get the opposite effect as the luminosity
decreases with increasing $B/B_{\rm eq}$. That happens because the source function, $S_{\nu}=\frac{P_{\nu}}{4\pi \alpha_{\nu}}$, where $P_{\nu}$ is the emission coefficient (eq.\ref{eq:P_nu_RL}), and $\alpha_{\nu}$ is the absorption coefficient (eq.\ref{eq:abs}), is $\propto B^{-1/2}$. Some of the disc
models lead to spectra which are optically thick ($\nu^{2.5}$) at $\nu>1$GHz. 
Such optically thick synchrotron emission is not observed in RQ AGN, even on mas scales. \citep{barvainis_2005,kukula_98,ulvestad_05}.
Thus, according to our model, the synchrotron source needs to extend up to $R_{\rm max}>3$~pc to push
the spectral turnover to lower frequencies. 
In contrast, most spherical corona models presented here 
become optically thick at frequencies well below 1~GHz for the assumed $R_{\rm max}=3$~pc.

    \begin{figure*}
    	\includegraphics[width=175mm]{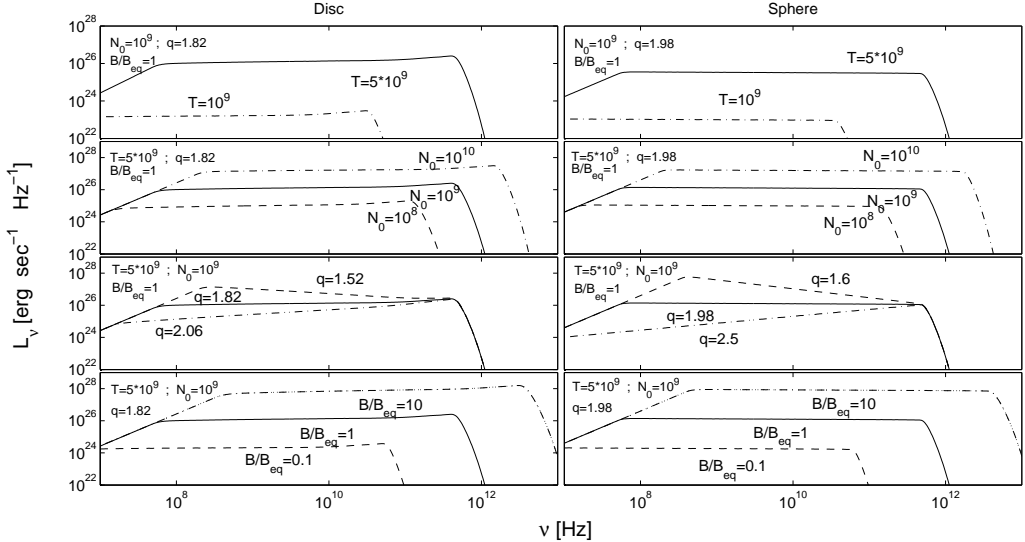}
    	\caption{The spectra of thermal electrons confined within a disc (left panels) and spherical (right panels) coronae. The other model parameters are the same as for the PL electrons (see caption of 
Fig.\ref{fig:pl_parameters} ).
The upper panel show the sharp dependence of the luminosity on $T$, as expected since analytic considerations
give $L_{\nu}\propto T^{3.84}$ (see text).
Note the overlap of the models in all panels, at the same $T$, in the optically thick part, where the emission becomes a
blackbody. Note also the sharp spectral break at high energies, when the whole configuration becomes optically thin,
in contrast with the $-1/2$ slope of the optically thin PL emission. The emission in the flat part of the continuum
shows similar dependence on the parameters explored, to the one shown by the PL electrons (Fig.\ref{fig:pl_parameters}).
 }
    	
    	\label{fig:thermal_parameters}
    \end{figure*}

Figure \ref{fig:thermal_parameters} explores the dependence of thermal synchrotron emission emission on various model parameters, as done in Fig.\ref{fig:pl_parameters} for the PL electrons. Here, however, we explore the dependence on $N$ and $T$ separately, and not just on $NkT$, as the synchrotron is produced by the thermal plasma, in contrast with the PL electrons where the thermal
plasma is used only to set the equipartition energy density of the PL electrons. 
The thermal plasma luminosity is significantly weaker than the PL luminosity, for the same parameters, as found for coronally
active stars \citep{gudel_2002,massi_92}.
Another general difference from the PL emission is the sharp drop at $\nu \sim 10^{11}-10^{13}$~Hz of the thermal synchrotron, in contrast with the
transition to optically thin synchrotron with a slope of $-1/2$.

The upper panel in Fig.8 shows the sharp dependence of the thermal synchrotron luminosity on $T$, where $L_{\nu}$ increases by
a factor of 725 when $T$ increases by a factor of 5 to $5\times 10^9$~K. This is expected since 
$\nu_{\rm peak}\propto T^{1.42}$ (eq.\ref{eq:our_nupeak_thermal}), while $L_{\nu}\propto
\nu_{\rm peak}^2T$ (eq.\ref{eq:Lnu_thermal_1}), which gives $L_{\nu}\propto T^{3.84}$. The numerical solution presented here 
yields $L_{\nu}\propto T^{4.09}$. 

At low frequencies, where the emission is optically thick, 
the spectrum is independent of $N$, $q$, and $B/B_{\rm eq}$, and depends only on $T$, as expected since 
the emission becomes a blackbody emission. The dependence 
of the emission in the flat part of the spectrum on $q, N_0$, and $B/B_{\rm eq}$,
is similar to the one presented by the PL electrons, as discussed above.

\subsection{The dependence on the PL electrons energy slope} \label{sec:radio_parameters_pl}
 
The value of $\delta = 2$, used above for the PL electron energy distribution, is motivated by the Fermi acceleration mechanism, and is observed
in various systems. Steeper values are expected when additional electron cooling processes
are taken into account (Longair 1994). The optically thin emission for $\delta = 2$ 
is $\nu ^
{-\frac{\delta -1 }{2}} $. However, steeper spectra are commonly observed in RQ AGN \citep{kukula_98,ulvestad_05,barvainis_2005,Behar_2015}, which imply larger values of $\delta$ are present. 

\begin{figure}
 	\includegraphics[width=80mm]{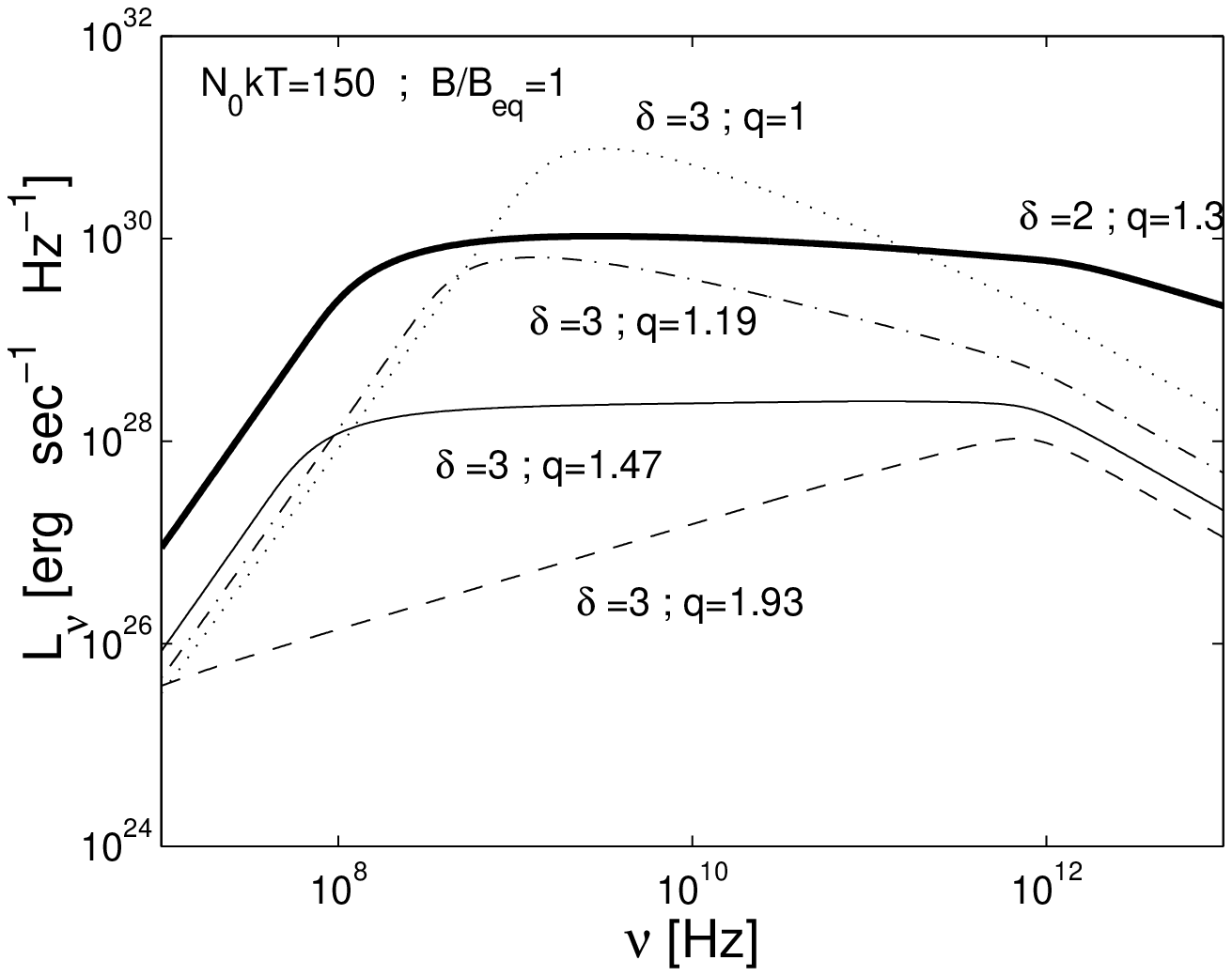}
 	\caption{A comparison of the spectra of disc coronae for electron PL distributions with
$\delta = 2$ and $\delta = 3$. As expected, the spectra in the $\delta = 3$ case are steeper.
The $q$ values required to derive spectral slopes of approximately $-1 , -1/2 , 0$ and $ 1/2$
are marked near each curve. Note the factor $\sim 100$ drop in the flat $L_{\nu}$ for the $\delta = 3$ case,
compared to the $\delta = 2$ case (see text). }
 	
 	\label{fig:pl_3}
 \end{figure}

Figure \ref{fig:pl_3} compares some of the earlier spectra with the spectra derived for the $\delta = 3$
case. As expected,
the spectra are steeper, keeping the other model parameters fixed. Alternatively, higher $q$ values
are required to derive a flat spectral slopes, as can also be seen from the analytical derivation
above (eq.\ref{eq:slope_disc_pl3}). A major difference is the drop in $L_{\nu}$ by about
a factor of 100, for models with a similar spectral slope. This results from the fact that 
the integrated electron energy in the $\delta=3$ case is concentrated near $\gamma_{\rm min}$, rather
than being evenly spread between $\gamma_{\rm min}$ and $\gamma_{\rm max}$ per logarithmic bin in $\gamma$, 
which is the case for $\delta=2$.

\subsection{The effect of a larger $R_{\rm in}$ } \label{sec:radio_parameters_Rin}

The inner disc boundary assumed above is $R_{\rm in}=3R_{\rm S}$, as expected for a viscous accretion disc
around a Schwarzschild black hole. However, studies of the observed UV spectral energy distribution of 
AGN, together with theoretical arguments, suggest the thin disc solution my break at a few 10’s of $R_{\rm S}$
\citep{laordavis_2014}, below which the accretion flow may change its nature (e.g. become a low radiative efficiency
geometrically thick flow). If the corona is formed only above the surface of a thin disc, it may not extend
down to $3R_{\rm S}$. For the specific model parameters used here, $M_{\rm BH}=10^8M_{\odot}$ shining at the
Eddington luminosity, we get that $R_{\rm in}=17R_{\rm S}$ from the analytic solution in \cite{laordavis_2014}.

\begin{figure}
    	\includegraphics[width=80mm]{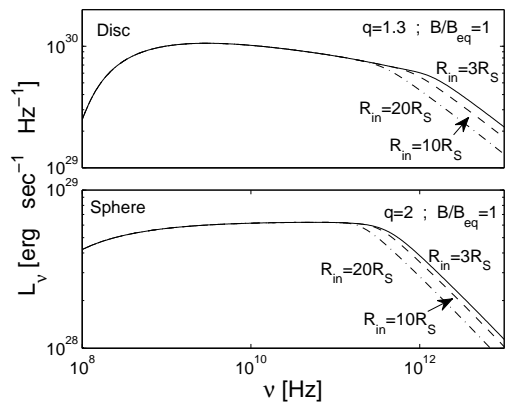}
    	\caption{The effect of increased $R_{\rm in}$ on the observed spectral shape from a PL electron
distribution. The values of
$R_{\rm in}$ are noted near each curve for the disc and sphere geometries. As expected, the break frequency
from a flat PL emission, to the steeper optically thin emission, decreases with increasing $R_{\rm in}$.
The spectral range at $\nu>3\times 10^{11}$~Hz is heavily dominated by dust emission, 
but it should be excluded by VLBI observations, which measure only the $T_b>10^8$~K component. 
This spectral turnover can be used to probe the coronal $R_{\rm in}$.}    	
    	\label{fig:inner_boundary}
    \end{figure}

Figure \ref{fig:inner_boundary} shows the effect of increasing  $R_{\rm in}$ on the observed spectrum from a disc corona of PL electrons. For the sake of completeness, the figure also shows the effect of a larger $R_{\rm in}$ for a spherical corona.
As expected, the frequency of the spectral break from a flat slope to an optically thin slope,
decreases with increasing  $R_{\rm in}$, from $\sim 1$~THz to a few hundred GHz, as the highest 
emission frequency originates from the smallest radii (see Fig.\ref{fig:inner_boundary}). One can also estimate the change
in the break frequency by applying the analytic estimate for $\nu_{\rm peak}$ for the innermost disc
parameters. One can use this break frequency to deduce the properties of the innermost 
corona. However, the spectral range above $\sim 300$~GHz is likely dominated by the Rayleigh-Jeans tail
of the coldest dust emission \citep{hughes_1993,haas_2000,haas_2013}, 
which rises extremely steeply and heavily dominates the emission above $\sim 300$~GHz. The dust emission may be overcome by VLBI observations, which
exclude sources with $T_b<10^8$~K, and may be able to overcome the strong background dust emission, and detect the expected synchrotron turnover in the emission of a compact
mas size source at the centre.

\section{Discussion}

The PL X-ray emission in AGN indicates the presence of coronal gas close to the centre. Since coronal emission in stars
is associated with synchrotron radio emission, this raises the possibility that the radio emission in RQ AGN is
also of coronal origin, in particular since RQ AGN follow the radio versus X-ray luminosity relation
displayed by stellar coronal emission (Gudel \& Benz 1993; LB08). 
The synchrotron emission can be produced by the coronal
hot thermal electrons, which also produce the X-ray emission. Or, it may be produced by PL electrons 
within the coronal gas. These PL electrons may be produced by magnetic
activity in the corona, specifically magnetic reconnection events, as seen in stellar coronae. 
The radio synchrotron emission then serves as an indicator for the coronal heating, while the X-rays
give the total coronal cooling (free-free in stellar coronae, and Comptonization in AGN coronae),
leading to the observed relation between the two. The PL electrons can also be produced by shock
acceleration, rather than reconnection, and may still serve as an indicator for the coronal heating,
if it is due to shocks.  Alternatively, the radio and X-ray relation may result from the fact that
both are produced by the same thermal electrons, which reside in a medium with a given ratio of
magnetic to radiation energy density.

Here we provide analytic estimates for the spectral slope and luminosity produced by thermal and PL electrons, 
embedded in a magnetic field in either a disc or a spherical configuration. We
present numerical solutions, and explore the solution parameter space, which can produce the observed
$L_{\nu}\sim 10^{28}-10^{30}$erg~s$^{-1}$~Hz$^{-1}$ at $\nu\sim 1-100$~GHz, which characterizes RQ AGN at
a bolometric luminosity of $10^{44}-10^{46}$erg~s$^{-1}$ (e.g. LB08). We concentrate here on models which
yield flat spectra, as this is the signatures of compact radio sources, such as AGN coronae.

Based on the typical rapid X-ray variability, the X-ray corona must be compact.
In contrast, the GHz radio corona must be on a pc scale. The models explored here allow to relate
the GHz emission to the emission at few hundred GHz, which is expected to originate from the X-ray corona.
As discussed in paper II, the much larger radio corona does not produce significant X-ray emission,
given the weakness of the disc flux on large scales,  which is the source for the Compton cooling X-ray emission of the corona.

\subsection{Comparison with observations}

Do the observed spectral slope in the radio indeed indicate significant emission from a compact
source? Are the spectra steep, indicating an optically thin synchrotron source with a size
of a pc or larger, or are the spectra flat, indicating the dominance of a compact
optically thick source? \cite{barvainis_96} find in 39 RQ AGN that the spectral shapes 
cover a wide range, with 40\% of the objects showing a flat or even rising spectral slopes.
Similar results are obtained in follow-up studies. \cite{kukula_98} find that 46\% (11/24)
of RQ AGN have $\eta>-0.2$, and 54\% $\eta<-0.6$, at 4.8-8.4~GHz, and \cite{barvainis_2005}
find 45\% (5/11) have $\eta>0.19$ and 55\% $\eta<-0.52$, also at the same frequency range.
Interestingly, with the small statistics available, the spectral slope distribution appears
to cluster at either steep or flat slopes, for both type 1 and type 2 AGN 
(Ulvestad \& Ho 2001; Lal \& Ho 2010). 
This suggests that the more extended optically thin
source, and the compact optically thick source do not go together, they are either unrelated,
or anti correlated. 

As noted above, significant unresolved pc scale radio emission is also derived from VLBI observations
\citep{blundell_98,caccianiga_2001,middelberg_2004,ulvestad_05,ulvestad_2005b,giroletti_2009,doi_2013,panessa_2013}.
In the case of NGC~1068, the pc scale emission is spatially resolved as coming from
a disc structure, but significant emission arises from structures on larger scales \citep{gallimore_97,gallimore_2004}. The spectrum of the pc scale emission is generally flat,
while the extended emission is steep, which indicates the high frequency emission is
likely dominated by the compact emission source. 

As noted by Barvainis et al. (1996; 2005), a significant fraction of the flat spectra sources
show variability, while none of the steep spectra sources varies on a timescale of months. 
A result consistent with 
the synchrotron emission models, where the flat sources need to be compact. The size
of the 5~GHz emitting region is similar to the size of the Broad Line Region, of $0.1L_{46}^{1/2}$~pc,
where $L_{46}=L_{\rm bol}/10^{46}~{\rm erg}~s^{-1}$ (LB08, eq.22), allowing significant
variability on the observed months timescale.  
  
Observations therefore clearly indicate that the coronal models, which predict a compact
emission region, may be valid in a significant fraction of AGN. The more extended emission
may also be powered by magnetized plasma ejected from the central compact source, however such
an extended component is not considered here.

\subsection{Comparison with earlier models}

A common interpretation for the radio emission in RQ AGN is a scaled down jet, compared
to the jet in RL AGN \citep{falke_95}. Apart from one possible exception \citep{blundell_2003},
VLBI monitoring of nearby RQ AGN detect either a static or slowly moving
radio sources \citep{ulvestad_05,reynolds_2009}, which argues against relativistic jets. 
A weak, non relativistic,
small scale jet is basically a cloud of magnetized plasma, and differs from a corona only in terms of the 
geometry. Since the physical size may be pc or smaller for both coronae and weak jets, 
it may remain unresolved, and therefore VLBI
imaging may not be able to differentiate the two. 

Since there are no robust prediction for the physical properties of a magnetized 
plasma in a weak jet, versus a corona, one cannot currently differentiate the two models
just based on the predicted synchrotron emission. 

As mentioned earlier, radio emission may also be produced by the interaction of an AGN driven
wind with the host galaxy interstellar medium. The radio emitting shock fronts are expected
to be on tens to a hundred pc scales \citep{jiang_2010}, and possibly out to kpc scale 
\citep{nims_2015}. In contrast, the spherical coronal models explored here are much more
compact, and produce the observed luminosity on a scale of a single pc. Spatially, the wind
interaction region may be resolved on VLBI mas scale imaging, which should show emission from
the shock front surface. In contrast, a magnetized CME may produce synchrotron from the whole volume.
However, the later case will be resolved only in nearby AGN (closer than 200~Mpc, for a 1~mas
resolution). 

A major difference, which results from the different physical sizes, is the
spectral slope. The size in the wind scenario suggests that the synchrotron emission will
be optically thick only at frequencies well below a GHz (LB08), and will therefore be optically
thin with $\eta\le -0.5$ at 1-100~GHz (unless the emission is highly clumped). 
In the CME case, $\eta> -0.5$ is possible, where the exact value depends on the radial
density and temperature distributions (see eqs.63, 66). As noted above, about half of the
RQ AGN are characterized by $\eta> -0.3$, which argues strongly against the wind scenario
in these objects.

If the wind scenarios apply, the observed radio emission should not vary on timescales of
a few years or shorter, while the compact coronal models allow variability on timescales
of a year and faster. As noted above, the observations suggest significant variability on 
timescales of months and below in some of the RQ AGN, which clearly rules out
the wind scenario as the dominant mechanism in these objects. 

Furthermore, if the spectral slope is flat, then the observed emission is a superposition 
of optically thick sources with a range of sizes, and the variability timescale is expected 
to decrease with increasing frequency, as possibly observed in NGC~7469 \citep{baldi_2015}.
 
\subsection{Thermal Synchrotron}

Is the thermal synchrotron a viable solution?
Here we find that the thermal synchrotron from an isothermal corona with 
$T\simeq 5\times 10^9$~K, as suggested by recent 
hard X-ray observations \citep{fabian_2015}, is a factor of $\sim 10^4$
lower than the emission of PL electrons, for similar model parameters (see Fig.8).
Thermal synchrotron can reach the observed luminosities assuming a higher temperatures. 
Since the thermal synchrotron follows $L_{\nu_{\rm peak}}\propto T^{4.09}$ (section \ref{sec:radio_parameters}), 
a value of $T\simeq 5\times 10^{10}$~K is required for the thermal synchrotron to reach
the synchrotron emissivity. This result can be understood analytically by comparing the
synchrotron source function $S_{\nu}=2.9\times 10^{-31}B^{-1/2}\nu^{5/2}~\rm{erg} ~\rm {cm}^{-2}  \rm{Hz}^{-1}  \rm{Strd} ^{-1}$ (eq.17 in LB08 for
$\delta=2$), to the blackbody source function $B_{\nu}=3.1\times 10^{-37}\nu^2T~ \rm{erg} ~\rm {cm}^{-2} \rm{Hz}^{-1}  \rm{Strd} ^{-1}$ in the Rayleigh-Jeans limit. This implies one needs $T\sim 10^6 \nu^{1/2} B^{-1/2}$~K, or
$T\sim 10^{11}$~K (for $\nu=10^{10} \rm{Hz}$, $B=1~\rm {Gauss}$) for the thermal 
synchrotron to become comparable to the synchrotron emission. The required $T$ can also be
estimated by equating the energy of the electrons emitting at a given frequency, 
$\nu=4.1\gamma^2 B$~MHz, to $kT$. Such a high value for $T$
appears to be excluded by X-ray observations (though these apply to the X-ray corona,
rather than the much larger radio corona at a few GHz). Such a high temperature is not expected from magnetic 
reconnection, and may also be excluded by pair production arguments, as discussed above
(section 1).

Alternatively, thermal synchrotron from a $T\simeq 5\times 10^9$~K corona may still be viable,
if the emitting surface is significantly larger. As noted above, current VLBI observations
lead to minimal brightness temperatures $\sim 10^9$~K, and therefore
do not exclude such thermal synchrotron. 

The signature of thermal synchrotron is an exponential cutoff in the emission above a few
hundred GHz, where the most compact emitting region becomes optically thin (eq.36). This results
from the exponential cutoff in the maximal electron energy density in the relativistic
Maxwell Boltzmann distribution (eq.14). This is in contrast with the PL distribution, where
the a maximal $\gamma=3000$ assumed here, ensures the peak emission is beyond 1000~GHz (section 3).
If there is a break in the PL electron energy distribution, it will also be associated by 
a corresponding spectral break at $\nu=4.1\gamma^2 B$~MHz (in the optically thin case).
However, the break will be to a steeper PL emission, rather than the sharp exponential break
of the thermal synchrotron.

\subsection{Coronal properties}

Is a thin corona at $T\simeq 5\times 10^9$K, which extends out to say $3\times 10^5R_{\rm S}$,
(3~pc in our model) a viable solution? This temperature is likely well above the escape speed 
from the disc surface at such a large radius. Therefore, the corona needs to be magnetically confined
to avoid a thermal wind with a considerable mass loss. The assumed equipartition magnetic field strength is
consistent with the value required by the magnetic confinement assumption. 

Since the system is powered by accretion, can accretion provide enough energy to power the 
radio emission? The fraction of the
rest mass energy dissipated in accretion from infinity to $R$ is $\epsilon= R_{\rm S}/R$ 
in the Newtonian limit (and a factor 3 larger in viscous accretion disc). At the outer
radius therefore $\epsilon\sim 10^{-5}$. Since the bolometric luminosity $L_{\rm bol}$ is likely derived from
$\epsilon\sim 0.1$, we expect that accretion can provide $L_{\rm radio}/L_{\rm bol}\sim 10^{-4}$,
which is well above the observed relation of $L_{\rm radio}/L_{\rm bol}\sim 10^{-6}$
(using $L_{\rm radio}/L_{\rm X-ray}\sim 10^{-5}$, and $L_{\rm X-ray}/L_{\rm bol}\sim 0.1$).
 
Clearly, the disc at such a large radius is self-gravitating and rather cold. Whether it can indeed maintain
a thin magnetically confined hot corona at its surface is an open question. 

Alternatively, and maybe more plausibly, the corona is in a spherical configuration. Although the solution here
is assumed static, such a configuration is likely formed by an outflow, i.e. a CME produced 
by coronal activity in the inner disc. In this case there is no need to confine the coronal gas,
and its power may come from the energy embedded in it when it is ejected. Interestingly, a flat spectral
slope is obtained for $q=2$, as expected in a uniform velocity outflow (the likely wind solution at
large radius). The coronal energy density required to get the observed 
$L_{\nu}\sim 10^{30}$~erg~s$^{-1}$~Hz$^{-1}$ is $u_{\rm corona}\sim 1400$~erg~cm$^{-3}$ (Fig.7, upper right panel)
at $R_0$. Since it scales at $R^{-2}$, the ratio to the radiation energy density of the AGN
$u_{\rm radiation}=L_{\rm bol}/4\pi R^2c$ remains constant. For $L_{\rm bol}=10^{46}$~erg~s$^{-1}$, which corresponds to the above $L_{\nu}$ at $\nu\sim 10$~GHz, one gets  
$u_{\rm corona}/u_{\rm radiation}=4\times 10^{-4}$. The corona is therefore a dynamically negligible
component. This constant ratio is interesting, and may provide some hints for the coronal heating mechanism.

\section{Conclusions}

We explore the possible radio emission from either a flat or a spherical magnetized corona, 
powered by either PL or thermal electrons. We concentrate on flat spectra models, which characterize 
about half of RQ AGN, and is the main signature of a compact emission region such as a corona.
Our main conclusions are as follows:

1. A flat spectral slope in the 1-1000 GHz range requires the superposition of synchrotron emission
from the innermost region at $3R_{\rm S}$ to $3\times 10^5R_{\rm S}$ ($3\times 10^{-5}$~pc to 3~pc).
The radio corona at 1~GHz emission comes from the largest scales, and 
should be resolved in mas resolution VLBI observations
of nearby AGN (closer than $\sim100$~Mpc), The few 100~GHz emission corona overlaps in
size the X-ray corona, and its size can only be constrained from its variability timescale.

2. The synchrotron emission at a given frequency, is produced over a wide range of 
radii for PL electrons. In contrast, the synchrotron emission of thermal 
electrons at a given frequency, originates from a narrow range of radii.

3. The synchrotron emission from a disc corona of PL electrons is nearly isotropic, as most of the
emission originates from the optically thin outer regions. The emission of thermal electrons 
from a disc corona shows a $\cos(\theta)$ dependence, as the observed emission is dominated by
the outermost optically thick region.

4. A luminosity of $L_{\nu}\sim 10^{30}$~erg~s$^{-1}$~Hz$^{-1}$ can be produced by 
PL electrons, magnetic field, and a corona, which are all in equipartition,
with an energy density which scales roughly as $R^{-1}$, and $NkT\sim 1000$~erg~cm$^{-3}$ at $3R_{\rm S}$. 
In the spherical corona configuration, the
energy density scales as $R^{-2}$. For PL electrons synchrotron, this 
equipartition energy density is a constant 
fraction of $\sim 4\times 10^{-4}$ of the central source radiation density.

5. Thermal synchrotron from $T\simeq 5\times 10^9$~K electrons, as suggested by recent 
hard X-ray observations, under predicts the radio  $L_{\nu}$ by a factor of $\sim 10^4$. To be a valid
mechanism the emitting surface must be significantly larger than assumed here, but it is not yet in 
contradiction with current
VLBI observations. Alternatively, one needs $T\simeq 5\times 10^{10}$~K to derive the observed
luminosity at $\nu\sim 10$~GHz.

6. At  $\nu =300$-1000~GHz the innermost corona is expected to become optically thin, and
the spectrum is expected to show a spectral break. The position of this break can be used
to probe the innermost coronal size. It should display a sharp cutoff, rather than a steeper PL, 
in case of thermal synchrotron emission. Since dust heavily dominates this spectral range, 
the detection requires VLBI observations, which exclude the low $T_b$ dust emission.

Additional constraints on the radio and X-ray coronal properties can be derived by including the
observed X-ray emission, in particular the enigmatic Gudel-Benz relation of $L_R/L_X\sim 10^{-5}$,
which is explored in paper II

Clearly, further exploration of the radio emission in RQ AGN, in particular at the mm range
\citep{Behar_2015}, will allow to probe directly the distribution of relativistic electrons
and magnetic fields closest to the centre. The relation of the mm emission with the X-ray emission, 
in particular their variabilities
\citep{baldi_2015}, can provide important insights for the physical mechanisms
which power accretion disc coronae in RQ AGN.

\section*{ACKNOWLEDGEMENTS}

We thank the referee, Nadia Zakamska, for the exceptionally thorough and helpful review. We also thank R. Sunyaev for suggesting to explore the coronal thermal synchrotron emission.
This research was supported by the Israel Science Foundation (grant no. 1561/13).

\end{document}